\documentclass[apj]{emulateapj} 
\usepackage{graphicx}
\usepackage{apjfonts}

\newcommand{\be}{\begin{equation}}
\newcommand{\ee}{\end{equation}}
\newcommand{\ba}{\begin{eqnarray}}
\newcommand{\ea}{\end{eqnarray}}

\newcommand{\mic}{\mu{\rm m}}

\newcommand{\kms}{\rm km\,s^{-1}}

\newcommand{\Mpc}{\rm Mpc}

\def\ls{\mathrel{\hbox{\rlap{\hbox{\lower4pt\hbox{$\sim$}}}\hbox{$<$}}}}
\def\gs{\mathrel{\hbox{\rlap{\hbox{\lower4pt\hbox{$\sim$}}}\hbox{$>$}}}}
\shorttitle{LoCuSS: X-ray AGN in local clusters}
\shortauthors{Haines et al.}
\begin{document}
\title{LoCuSS: A dynamical analysis of X-ray AGN in local clusters}

\author{
  C.\ P.\ Haines,$\!$\altaffilmark{1,2}
  M.\ J.\ Pereira,$\!$\altaffilmark{1}
  A.\ J.\ R.\ Sanderson,$\!$\altaffilmark{2}
  G.\ P.\ Smith,$\!$\altaffilmark{2}
  E.\ Egami,$\!$\altaffilmark{1}
  A.\ Babul,$\!$\altaffilmark{3}
  A.\ C.\ Edge,$\!$\altaffilmark{4}\\
  A.\ Finoguenov,$\!$\altaffilmark{5,6}
  S.\ M.\ Moran,$\!$\altaffilmark{7}
  N.\ Okabe$\!$\altaffilmark{8}\\
}

\altaffiltext{1}{Steward Observatory, University of Arizona, 933 North
 Cherry Avenue, Tucson, AZ 85721, USA; cphaines@as.arizona.edu} 
\altaffiltext{2}{School of Physics and Astronomy, University of
  Birmingham, Edgbaston, Birmingham, B15 2TT, UK}
\altaffiltext{3}{Department of Physics and Astronomy, University of Victoria, 3800 Finnerty Road, Victoria, BC, V8P 1A1, Canada}
\altaffiltext{4}{Institute for Computational Cosmology, Department of Physics, University of Durham, South Road, Durham, DH1 3LE, UK} 
\altaffiltext{5}{Max-Planck-Institut f\"{u}r extraterrestrische Physik, Giessenbachstra{\ss}e,85748 Garching, Germany}
\altaffiltext{6}{Center for Space Science Technology, University of Maryland Baltimore County, 1000 Hilltop Circle, Baltimore, MD 21250, USA}
\altaffiltext{7}{Department of Physics and Astronomy, The Johns
  Hopkins University, 3400 N. Charles Street, Baltimore, MD 21218,
  USA} 
\altaffiltext{8}{Academia Sinica Institute of Astronomy and Astrophysics (ASIAA), P.O. Box 23--141, Taipei 10617, Taiwan}


\begin{abstract}
We present a study of the distribution of X-ray AGN in a representative sample of 26 massive clusters at $0.15{<}z{<}0.30$, combining {\em Chandra} observations sensitive to X-ray point sources of luminosity $L_{X}{\sim}10^{42}$\,erg\,s$^{-1}$ at the cluster redshift with extensive and highly complete spectroscopy of cluster members down to ${\sim}{\rm M}^{*}_{K}{+}2$. In total we identify 48 X-ray AGN among the cluster members, with luminosities $2{\times}10^{41}-1{\times}10^{44}$\,erg\,s$^{-1}$. Based on these identifications, we estimate that $0.73{\pm}0.14$\% of cluster galaxies brighter than $M_{K}{=}{-}23.1$ ($M_{K}^{*}{+}1.5$) host an X-ray AGN with $L_{X}{>}10^{42}$\,erg\,s$^{-1}$. In the stacked caustic diagram that shows $(v_{los}{-}{<}v{>})/\sigma_{v}$ versus $r_{proj}/r_{500}$, the X-ray AGN appear to preferentially lie along the caustics, suggestive of an infalling population. 
They also appear to avoid the region with lowest cluster-centric radii and relative velocities ($r_{proj}{<}0.4r_{500}$; \mbox{$|v{-}{<}v{>}|/\sigma_{v}{<}0.8$}), which is dominated by the virialized population of galaxies accreted earliest into the clusters. 
The line-of-sight velocity histogram of the X-ray AGN shows a relatively flat distribution, and is inconsistent with the Gaussian distribution expected for a virialized population at 98.9\% confidence. 
Moreover the velocity dispersion of the 48 X-ray AGN is $1.51{\times}$ that of the overall cluster population, which is consistent with the $\sqrt{2}$ ratio expected by simple energetic arguments when comparing infalling versus virialized populations. This kinematic segregation is significant at the 4.66-$\sigma$ level. 
When splitting the X-ray AGN sample into two according to X-ray or infrared (IR) luminosity, both X-ray bright ($L_{X}{>}10^{42}$) and IR-bright ($L_{TIR}{>}2{\times}10^{10}L_{\odot}$) sub-samples show higher velocity dispersions than their X-ray dim and IR-dim counterparts at ${>}2{\sigma}$ significance. This is consistent with the nuclear activity responsible for the X-ray and IR emission being slowly shut down as the host galaxies are accreted into the cluster. 
Overall our results provide the strongest observational evidence to date that X-ray AGN found in massive clusters are an infalling population, and that the cluster environment very effectively suppresses radiatively-efficient nuclear activity in its member galaxies. These results are consistent with the view that for galaxies to host an X-ray AGN they should be the central galaxy within their dark matter halo and have a ready supply of cold gas.

\end{abstract}

\keywords{ galaxies: active --- galaxies: clusters: general ---
  galaxies: evolution --- galaxies: stellar content }

\section{Introduction}
\label{intro}

\setcounter{footnote}{7}

In recent years observations have shown that at the heart of most, if not all, massive galaxies is a supermassive black hole (SMBH) \citep[for a review see][]{ferrarese}. 
The observed tight correlations between the SMBH and galaxy bulge masses \citep[e.g.][]{haring} or stellar velocity dispersion \citep{gebhardt,ferrarese}, imply that the evolution of the galaxy and the central black hole are fundamentally intertwined. \citet{silk} suggested that these tight correlations arise naturally from the self-regulated growth of SMBHs through accretion of material as active galactic nuclei (AGN), triggered by the merger of gas-rich galaxies. Tidal torques produced by the merger channel large amounts of gas onto the central nucleus, fuelling a powerful starburst and rapid black hole growth, until feedback from radiatively efficient accretion is able to drive quasar winds and expel the remaining gas from the galaxy \citep[e.g.][]{barnes,silk,springel,hopkins08}.

There have been significant advances of late in the theoretical framework for understanding galaxy evolution by incorporating AGN feedback, in particular to reproduce the global properties of galaxies observed in the current large scale surveys such as the SDSS \citep[e.g.][]{bower,croton,hopkins06,schaye}. The enormous amounts of energy produced by AGN have also been invoked as efficient mechanisms to quench star-formation in galaxies, either via extreme feedback from quasars \citep{springel,hopkins,hopkins06,hopkins08}, or via low-level, quasi-continuous feedback from radiatively inefficient accretion of hot gas from the X-ray emitting halos of massive galaxies \citep{best,croton,hardcastle}. 

Given the fundamental link between SMBHs and bulge growth as well as the importance of AGN feedback for the evolution of the host galaxy, a key unresolved issue in astrophysics is determining the source of gas that fuels the growth of SMBHs and resultant nuclear activity, and what triggers the gas inflow. As well as the previously described merger paradigm, secular processes including bar-driven gas inflows \citep{kormendy,jogee}, disk-instabilities, stochastic collisions with molecular clouds \citep{hop+hern}, and stellar winds from evolved stars \citep{ciotti} have all been proposed as means to supply gas onto SMBHs and trigger their activity. 

A primary complication for understanding the nature of AGN and their impact on galaxy evolution is the observed zoo of AGN populations and classes. These are variously identified via optical spectroscopy or characteristic emission at X-ray, radio or infrared wavelengths, and each one shows a diverse range of Eddington ratios or host galaxy properties. 
In principle, X-ray emission provides the cleanest and most reliable approach to selecting AGN and in particular identifying those galaxies undergoing significant black hole growth. The hard X-ray emission is directly associated with accretion close in to the black hole (10--100 gravitational radii). It is produced in the hot corona that surrounds the black hole by the inverse Compton scattering of ultraviolet photons emitted by the accretion disk. Except for strongly absorbed AGN (N$_{H}{>}10^{23}$\,cm$^{-2}$) which are rare at low-redshifts \citep{tozzi,burlon}, X-ray selection is nearly independent of obscuration. In contrast the selection of AGN at UV/optical wavelengths is strongly affected by absorption and orientation effects, as well as being prone to contamination by stellar light. Moreover, while X-ray and infrared-selected AGN typically have Eddington ratios of $\sim$1--10\% \citep{hickox}, associating them with rapid black hole growth, radio-loud AGN and low-luminosity optically-selected AGN (LINERs) are characterized by much lower Eddington ratios (${<}10^{-3}$).

The environments, galaxy hosts and clustering of the diverse classes of AGN can provide important clues to understanding both the accretion processes powering the AGN (e.g. whether it is fuelled by the accretion of hot or cold gas), and its subsequent impact on the galaxy host in terms of building up its bulge or quenching its star-formation. This has motivated numerous authors who have examined the AGN populations in massive galaxy clusters \citep[e.g.][]{dressler,martini,martini07,martini09} where, for the virialized galaxy population at least, only hot gas is available for accretion onto the SMBHs. 

The distribution of AGN in galaxy clusters provides a fundamental test for those accretion processes which require a ready supply of cold gas in the host galaxy, such as secular bar/disk-instabilities \citep{hop+hern} or the merger of two gas-rich galaxies. Given the much lower fraction of galaxies with substantial reservoirs of H{\sc i} gas found in cluster cores compared to the field \citep[e.g.][]{giovanelli,cortese}, nuclear activity among the virialized cluster population should also be reduced. However, the molecular gas contents of late-type cluster galaxies appear normal \citep{boselli}, while \citet{young} and \citet{haines11} show that at least some cluster S0s are able to retain their molecular gas and dust contents through several Gyr while virialized within the cluster. As disk-instabilities are described by the Toomre criterion, it may be sufficient to simply {\em reduce} the gas surface density of cluster galaxies to prevent secular triggering of AGN, by making the remaining gas stable against collapse. 
In the case of the merger paradigm, nuclear activity should be even more strongly suppressed within rich clusters. Here the encounter velocities of galaxies are much greater than their internal velocity dispersions, preventing their coalescence, in spite of the high galaxy densities \citep{aarseth}. Gas-rich mergers should instead be most frequent in galaxy groups \citep{hopkins08} and the cluster outskirts, where many galaxies (including gas-rich ones), under the influence of the cluster's tidal field, are part of a convergent flow resulting in enhanced interactions between neighbors \citep{vandeweygaert}. Since these galaxies are  falling into the cluster for the first time, they have yet to have their gas contents be affected by their passage through the dense intra-cluster medium (ICM). Galaxy harassment due to frequent high-speed fly-by interactions has also been proposed as a means of triggering nuclear activity, by driving dynamical instabilities that efficiently channel gas onto the SMBHs \citep{moore}, although again this requires the host galaxy to contain a gas reservoir. 

The trends predicted by associating X-ray AGN to gas-rich galaxies could be diluted however, or even reversed, by the tendency of X-ray AGN (above a given $L_{X}$) to be hosted by the most luminous galaxies \citep{sivakoff,tasse}, a population which is most centrally concentrated within clusters \citep{lin04,thomas}. Both \citet{sivakoff} and \citet{haggard} show the X-ray AGN fraction to increase by an order of magnitude from low-mass populations ($\mathcal{M}{\sim}10^{10}{\rm M}_{\odot}$) to the most massive galaxies ($\mathcal{M}{\ga}10^{11}{\rm M}_{\odot}$) in both cluster and field populations. These trends act as physical selection effect, as the most luminous galaxies are more likely bulge-dominated and hence have higher black hole masses. Thus for the same accretion rate relative to the Eddington limit, the higher mass galaxy, will likely have a higher X-ray luminosity, and be more likely detected above a fixed $L_{X}$ limit.

If nuclear activity is not dependent on a cold gas supply, but is being instead fuelled by the accretion of hot gas from the halos of massive galaxies \citep{croton} or recycled gas from evolved stars \citep{ciotti} then we would expect nuclear activity to be most prevalent among the massive elliptical galaxies that dominate cluster cores. In this case the AGN fraction should be higher in clusters than the field.

The earliest works identified AGN via their optical emission lines \citep{gisler,dressler}, finding them to be much rarer among cluster members than in field samples. More recent studies based on the SDSS have obtained rather conflicting results, with some confirming the previous decline in the AGN fraction with galaxy density \citep{kauffmann04,mahajan,hwang}, while others find no apparent trend with environment \citep{miller,haines07}. \citet{vonderlinden} find that while overall the fraction of galaxies which host a powerful optical AGN (with $L$[O{\sc iii}$]{>}10^{7}L_{\odot}$) decreases towards the cores of clusters, when considering just star-forming galaxies, the AGN fraction remains constant with cluster-centric radius. They suggest that the decline in the AGN fraction is almost exclusively an effect of the increasing number of red sequence galaxies, a population which hosts very few strong AGN.  

The launches of the {\em Chandra} and {\em XMM X-ray Observatories} have opened up an efficient way of identifying X-ray AGN in and around massive clusters from the same images used to probe the fundamental properties of the clusters themselves (e.g. masses, $T_{ICM}$). 
Statistical analyses of X-ray point sources in cluster fields have reported overdensities of X-ray sources with respect to non-cluster fields \citep[e.g.][]{cappi,molnar,ruderman,branchesi}. \citet{gilmour} found an excess of ${\sim}1.5$ X-ray point sources per cluster within 1\,Mpc from a sample of 148 clusters at $0.1{<}z{<}0.9$. They found the radial distribution of these excess sources consistent with a flat radial distribution within 1\,Mpc, although they could also be consistent with being drawn from the general cluster galaxy population. \citet{ruderman} found instead a central spike within 0.5\,Mpc followed by a secondary broad peak at the virial radius for relaxed clusters, with flatter spatial distributions for disturbed systems. \citet{koulouridis} instead found that the overdensity of X-ray point sources in 16 clusters at $0.07{<}z{<}0.28$ to be a factor ${\sim}4$ less than that of bright optical galaxies, and concluded that the triggering of luminous ($L_{X}{>}10^{42}$\,erg\,s$^{-1}$) X-ray AGN to be strongly suppressed in rich clusters.

\citet{martini} performed a redshift survey of X-ray point sources in 8 clusters at $0.06{<}z{<}0.31$, identifying 30 X-ray AGN with $L_{X}{>}10^{41}$\,erg\,s$^{-1}$ and $M_{R}{\le}{-}20$. Six of these clusters were X-ray luminous systems with $L_{X}{>}4{\times}10^{44}$\,erg\,s$^{-1}$, including Abell 1689. The other two were the pair of low-mass merging clusters Abell 3125 and Abell 3128 at $z{=}0.06$. Abell 3125 in fact does not have a diffuse ICM, while the ICM for Abell 3128 is double peaked.
The resulting X-ray AGN fraction of ${\sim}5$\% among cluster galaxies brighter than $M_{R}{=}{-}20$, was much higher than previously expected and remarkably similar to those found in the field \citep{martini07,lehmer}. This suggested little or no environmental dependence and that cluster galaxies could retain significant reservoirs of cold gas near their central SMBHs. The X-ray AGN fraction was highest (${\sim}12$\%) in the two low-mass merging systems Abell 3125/3128.  
Only a small fraction of these X-ray AGN would have been classified as such from their optical spectra, and so the difference in the observed environmental trends seen in optically-selected QSOs in which they avoid cluster cores \citep[e.g.][]{dressler,sochting}, was put down to biases against detection of low-luminosity AGN due to host galaxy dilution and obscuration.
In the dynamical analysis of the same sample, both the stacked velocity and radial distributions of the X-ray AGN were found to be entirely consistent with being drawn from the overall cluster galaxy population \citep{martini07}. 
The most luminous X-ray AGN with $L_{X}{>}10^{42}$\,erg\,s$^{-1}$ were however found to be {\em more} centrally concentrated than inactive cluster galaxies. This study had little coverage of the cluster outskirts, with 90\% of the X-ray AGN within 0.5\,$r_{200}$. However, \citet{sivakoff} using mosaics of {\em Chandra} images covering the virialized regions of Abell 85 and Abell 754 (two merging clusters at $z{=}0.055$ with $L_{X}{\sim}5{\times}10^{44}$\,erg\,s$^{-1}$), obtained similar trends, with again no differences in the spatial distributions (both radially or in velocity) of X-ray AGN (with $L_{X}{>}10^{41}$\,erg\,s$^{-1}$) and non-active cluster galaxies. \citet{gilmour07} instead found that the 11 X-ray AGN within the supercluster A901/2 at $z{\sim}0.17$ lie predominately in intermediate density regions, avoiding the cluster cores, prefering instead the cluster outskirts or group-like environments containing more blue galaxies than on average. 

From these analyses it remains unclear whether the X-ray AGN observed in galaxy clusters should be considered to be virialized or an infalling population. Given the small numbers of clusters studied in detail with extensive spectroscopy, 
it is unlikely that the previous three samples can be regarded as representative of the general massive cluster population. The samples of \citet{martini}, \citet{sivakoff} and \citet{gilmour07} all appear overrepresented by merging systems. 
What is lacking is a statistical sample of spectroscopically-confirmed X-ray AGN from a large, representative set of massive clusters at a specific epoch. The aim of this present work is to produce such a sample, taking advantage of the unique multi-wavelength dataset recently assembled within LoCuSS (Local Cluster Substructure Survey)\footnote{www.sr.bham.ac.uk/locuss/}.

LoCuSS is a multi-wavelength survey of X-ray luminous galaxy clusters at $0.15{\le}z{\le}0.3$ drawn from the ROSAT All Sky Survey cluster catalogs \citep{ebeling98,ebeling00,bohringer}. LoCuSS comprises three broad inter-related themes: (i) calibration of mass-observable scaling relations for cluster cosmology, (ii) physics of cluster cores, and (iii) the physics of galaxy transformation in group/cluster environments. This is the latest in a series of papers in the third ``galaxy evolution'' theme. The science goals and design of this theme are described in detail by \citet{smith}. In summary, we have obtained panoramic multi-wavelength data (FUV--FIR) on a morphologically unbiased sample of 30 clusters whose X-ray luminosities span $3.79{\times}10^{44}{\le}L_{X}($0.1--2.4\,keV$){\le}2.28{\times}10^{45}$\,erg\,s$^{-1}$. These data span fields of ${\ga}25^{\prime}{\times}25^{\prime}$ -- i.e.\ out to ${\sim}1$.5--2 cluster virial radii \citep{haines09,pereira,smith}. These 30 clusters were selected from the parent sample simply on the basis of being observable by Subaru on the nights allocated to us \citep{okabe09}, and should therefore not suffer any gross biases towards (for example) cool core cluster, merging clusters etc. Indeed \citet{okabe09} show that the sample is statistically indistinguishable from a volume-limited sample. 

Twenty-six of the 30 clusters have been observed with {\em Chandra} to depths suitable for identifying X-ray AGN within the clusters. We have combined these X-ray data with our extensive spectroscopy (${\sim}2$00--400 members per cluster) from MMT/Hectospec, to derive a highly complete census of the X-ray AGN population within this statistical sample of local clusters. We have performed a dynamical analysis of these cluster X-ray AGN in order to understand whether they are hosted primarily by galaxies still infalling into the cluster (and which may have yet to encounter the dense ICM) or rather galaxies which were accreted at early epochs and are now virialized. This analysis is aided by comparison to predictions from the Millennium cosmological simulation of the distributions of galaxies in the stacked caustic diagrams coded according to the epoch at which they were accreted into the cluster. We also present the X-ray luminosity function of cluster AGN and examine the global optical and infrared properties of the galaxies hosting X-ray AGN in clusters. 

In \S\ref{sec:data} we summarize the data used in this paper, while our use of cosmological simulations is described in \S\ref{sec:dynamic}.  The main results are then presented in \S\ref{sec:results}, followed by a discussion and summary in \S\ref{sec:discuss} and \S\ref{sec:summary}.  Throughout we assume \mbox{$\Omega_M{=}0.3$}, \mbox{$\Omega_\Lambda{=}0.7$} and \mbox{${\rm H}_0{=}70\,\kms\Mpc^{-1}$}, for consistency with our previous work \citep{haines09}.

\section{Data}
\label{sec:data}

This sample of 26 clusters was chosen because the clusters all have publicly available {\em Chandra} ACIS data with exposure times of ${\sim}1$0\,ksec or greater, sufficient to detect X-ray sources as faint as $L_{X}{\sim}10^{42}$\,erg\,s$^{-1}$ at the cluster redshift, and to derive robust measures of $r_{500}$ from the extended X-ray emission from the ICM. Of the original sample of 30 clusters, A291 and A2345 were excluded due to lack of {\em Chandra} imaging, while it was not possible to determine $r_{500}$ values for A689 or Z348.
For each of the clusters we have panoramic near-IR imaging to $K{\sim}19$, $J{\sim}21$, obtained with either WFCAM on the 3.8-m United Kingdom Infrared Telescope (UKIRT) which covers $52^{\prime}{\times}52^{\prime}$ regions centered on each cluster (23 out of 26 clusters), or NEWFIRM on the 4.0-m telescope at Kitt Peak which provides a field of view of $27^{\prime}{\times}27^{\prime}$ for the remaining 3 systems \citep{haines09}. For each cluster we also have deep {\em Subaru} optical ($V,i$) imaging covering $34^{\prime}{\times}27^{\prime}$, used by \citet{okabe09} to perform weak lensing mass estimates. Additionally, $ugriz$ photometry to $r{\sim}22$ is available for 22 of the 26 clusters, taken from the eighth data release from the Sloan Digital Sky Survey \citep[SDSS-DR8;][]{aihara}.
Each cluster was observed across a $25^{\prime}{\times}25^{\prime}$ field of view at 24$\mu$m with MIPS \citep{rieke04} on board the {\em Spitzer Space Telescope}\footnote{This work is based in part on observations made with the Spitzer Space Telescope, which is operated by the Jet Propulsion Laboratory, California Institute of Technology under a contract with NASA (contract 1407)} \citep{werner} (PID: 40872; PI: G.~P. Smith), reaching typical 90\% completeness limits of 400$\mu$Jy, sensitive to obscured star formation within cluster galaxies to levels of 3\,M$_{\odot}$\,yr$^{-1}$ in our most distant systems. {\em Herschel} PACS/SPIRE imaging from our LoCuSS {\em Herschel} Key Programme provides 100--500$\mu$m photometry for  matched $25^{\prime}{\times}25^{\prime}$ fields \citep{smith}, reaching comparable sensitivities in terms of obscured SFRs.

\subsection{MMT/Hectospec spectroscopy}
\label{spectroscopy}

We have recently completed ACReS (the Arizona Cluster Redshift Survey; Pereira et al. in preparation) a long-term programme to observe our sample of 30 galaxy clusters with MMT/Hectospec. Hectospec is a 300-fiber multi-object spectrograph with a circular field of view of $1^{\circ}$ diameter \citep{fabricant} on the 6.5m MMT telescope at Mount Hopkins, Arizona. The details of the observations, targeting strategy and data reduction are presented elsewhere (Haines et al. 2009a,b; 2010; Pereira et al. in preparation). 

Briefly, probable cluster galaxies are targeted according to their $K$-band absolute magnitude and $J{-}K$ color, prioritizing those galaxies detected with {\em Spitzer}/MIPS at 24$\mu$m. The targeting is based on the empirical observation that galaxies of a particular redshift lie along a single narrow $J{-}K/K$ color-magnitude (C-M) relation, which evolves redward monotonically with redshift to $z{\sim}0.5$ \citep{haines09b}. The NIR colors of galaxies are relatively insensitive to star-formation history and dust extinction. This effectively allows the creation of a stellar mass-limited sample within a narrow redshift slice centered on the cluster, with no preference towards passive (``red sequence'') or star-forming (``blue cloud'') sub-populations, simply by selecting galaxies within a color slice of width 0.3--0.4\,mag enclosing the observed $J{-}K/K$ C-M relation for that cluster \citep{haines09}. 
For each cluster, galaxies brighter than $M_{K}^{*}{+}2.0$ were targeted (falling to $M_{K}^{*}{+}1.5$ for particularly rich systems at higher redshifts) over the full UKIRT/WFCAM field, reaching overall completeness levels (as measured within $r_{200}$) of 70\% for the M$_{K}$-selected sample, increasing to 96.4\% (5245/5441) for those galaxies detected with {\em Spitzer}. 

The selection of galaxy targets was made without any prior knowledge of whether they were detected with {\em Chandra} or not, and so should not have any bias with respect to X-ray AGN. However, X-ray AGN are more likely to be MIR-bright than normal galaxies of the same stellar mass \citep{treister}, and so are more likely to have been targeted. This was accounted for explicitly by weighting each galaxy according to the probability that it was targeted for spectroscopy based on its $M_{K}$, 24$\mu$m flux and spatial location within the cluster. Moreover, our spectroscopic survey should have no {\em morphological} bias against unresolved extra-galactic objects such as compact galaxies or quasars. This is because we targeted sources based solely upon their near-IR colors or 24$\mu$m emission, irrespective of whether they are resolved or not in our near-IR images. Stars show much bluer near-IR colors ($J{-}K{<}1$) than galaxies or quasars at the redshift of interest \citep[see Figs 2, 3 of][]{haines09b}, and were identified and excluded as targets for spectroscopy {\em only} if they have $J{-}K{<}1$ and are unresolved in our $K$-band data.

\begin{table}
\caption{Summary of Chandra observations.}
\begin{center}
\begin{tabular}{lcccrcc} \hline
Cluster & $z$ & $r_{500}$ & {\em Chandra} & Exp time & $L_{X,lim}$ & $N_{AGN}$ \\ 
Name & & (Mpc) & ID (Inst) & (ksec) & \hspace{-0.1cm}($10^{41}$erg/s)\hspace{-0.2cm} & \\ \hline
Abell 68    & 0.2510 & 0.955 & 3250 (I) & 10.0 & 10.4 & 0\\
Abell 115   & 0.1919 & 1.304 & 3233 (I) & 49.7 & 1.16 & 5\\
Abell 209   & 0.2092 & 1.230 & 3579 (I) & 10.0 & 6.99 & 3\\
Abell 267   & 0.2275 & 0.994 & 3580 (I) & 19.9 & 4.22 & 1\\
Abell 383   & 0.1887 & 1.049 & 2320 (I) & 19.3 & 2.88 & 2\\
Abell 586   & 0.1707 & 1.150 & 530 (I) & 10.0 & 4.47 & 2\\
Abell 611   & 0.2864 & 1.372 & 3194 (S) & 35.1 & 3.88 & 5\\
Abell 665   & 0.1827 & 1.381 & 3586 (I) & 29.7 & 1.75 & 3\\
Abell 697   & 0.2818 & 1.505 & 4217 (I) & 19.5 & 6.91 & 0\\
Abell 963   & 0.2043 & 1.275 & 903 (S) & 36.3 & 1.83 & 1\\
Abell 1689  & 0.1851 & 1.501 & 5004 (I) & 19.9 & 2.71 & 4\\
Abell 1758  & 0.2775 & 1.376 & 2213 (S) & 58.3 & 2.26 & 1\\
Abell 1763  & 0.2323 & 1.220 & 3591 (I) & 19.6 & 4.48 & 2\\
Abell 1835  & 0.2520 & 1.589 & 6880 (I) &117.9 & 0.89 & 6\\
Abell 1914  & 0.1671 & 1.560 & 3593 (I) & 18.9 & 2.28 & 1\\
Abell 2218  & 0.1733 & 1.258 & 1666 (I) & 41.2 & 0.96 & 3\\
Abell 2219  & 0.2257 & 1.494 & 896 (S) & 42.3 & 1.95 & 2\\
Abell 2390  & 0.2291 & 1.503 & 4193 (S) & 76.8 & 0.90 & 5\\
Abell 2485  & 0.2476 & 0.830 & 10439 (I) & 19.8 & 5.09 & 1\\
RXJ0142.0+2131 & 0.2771 & 1.136 & 10440 (I) & 19.9 & 6.52 & 0\\
RXJ1720.1+2638 & 0.1599 & 1.530 & 4361 (I) & 24.0 & 1.52 & 0\\
RXJ2129.6+0005 & 0.2337 & 1.227 & 552 (I) & 10.0 & 8.94 & 0\\
Z\,1693   & 0.2261 & 1.050 & 10441 (I) & 21.4 & 3.87 & 0\\
Z\,1883   & 0.1931 & 1.107 & 2224 (S) & 25.7 & 1.97 & 1\\
Z\,2089   & 0.2344 & 1.024 & 7897 (I) &  9.0 & 9.90 & 0\\
Z\,7160   & 0.2565 & 1.128 & 4192 (I) & 91.6 & 1.19 & 0\\ \hline
\end{tabular}
\end{center}
\label{chandra}
\end{table}

To date, ACReS has required the equivalent of 13 full nights of observations since December 2008, producing ${\sim}3$0\,000 spectra, of which ${\sim}1$0\,000 have been identified as being cluster members, the largest sample to date from a redshift survey targeting galaxy clusters. We have obtained 3--6 fiber configurations per cluster resulting in typically redshifts for 100--500 cluster members, the number depending primarily on the richness and/or compactness of the cluster. The 270 line grating was used, providing coverage over the wavelength range 3650--9200{\AA} at 6.2{\AA} resolution. For each cluster, we plotted redshift against cluster-centric radius, identifying cluster members as lying within the general ``trumpet''-shaped caustic profile expected for galaxies infalling and orbiting within a massive structure (see {\S}\ref{sec:dynamic}). 

\subsection{Chandra X-ray imaging}
\label{dataxray}

The 26 clusters have available {\em Chandra} data, both from the archive and drawn from our own {\em Chandra} Cycle 10 programme (PID: 10800565). The observations of 20/26 clusters were made with the I mode of the Advanced Camera for Imaging Spectroscopy (ACIS-I), the remaining six being observed with ACIS-S. The exposure times were typically 20ksec, but range between 9--120ksec. The larger ACIS-I field of view is $16.9^{\prime}{\times}16.9^{\prime}$, slightly smaller than our {\em Spitzer}/MIPS fields, meaning that all sources detected by {\em Chandra} are also covered by our 24$\mu$m imaging. The ACIS-I instrument is made up of 4 CCDs in a $2{\times}2$ grid. The cluster core is always centered within one of these CCDs to minimize the impact of the gaps between the CCDs on measurements of the ICM, and hence the spatial coverage of the clusters is rather asymmetric. The ACIS-S imaging provides a smaller field of view of $8.3^{\prime}{\times}8.3^{\prime}$ from a single CCD, centered on the cluster core. 

The deprojected dark matter densities, gas densities and gas temperature profiles were derived by fitting the phenomenological cluster models of \citet{ascasibar} to a series of annular spectra extracted for each cluster \citep{sanderson10}. The best-fitting cluster models were then used to estimate $r_{500}$, the radius enclosing a mean overdensity of 500 with respect to the critical density of the Universe at the cluster redshift \citep{sanderson09}. A summary of the clusters observed, redshifts and $r_{500}$ values, Chandra observation ID and ACIS instrument used and total exposure times after cleaning is shown in Table~\ref{chandra}. Details of the reductions and analysis can be found in \citet{sanderson09}. 

\begin{figure}
\plotone{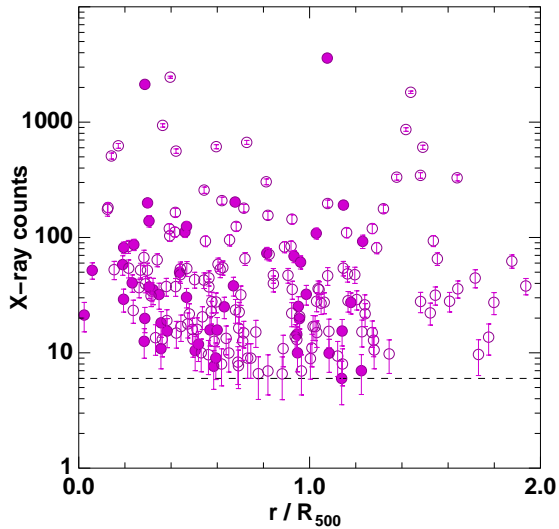}
\caption{Broadband (0.3--7\,keV) X-ray counts of detected point sources with known redshifts as a function of projected clustercentric radius, revealing the impact of the extended X-ray emission from the ICM on point source detection. Solid (open) symbols indicate X-ray point sources associated to cluster members (non-members). The horizontal dashed line indicates the minimum requirement for detection of 6 counts.} 
\label{xrayradial}
\end{figure}

To detect X-ray point-sources that are potential X-ray AGN, the wavelet-detection algorithm {\sc ciao wavdetect} was applied, requiring a minimum of six counts in the broad energy (0.3--7\,keV) range to be considered in this analysis.
The observed fluxes in this band were derived assuming a $\Gamma{=}1.7$ power-law spectrum with Galactic absorption, following \citet{kenter}. We then calculated the rest-frame luminosity for all sources with redshifts, assuming k-corrections of the form $(1+z)^{\Gamma-2}$. 

For our typical exposure time of 20\,ksec, our survey sensitivity limit of six counts corresponds to an X-ray luminosity of $L_{X}{=}3.5{\times}10^{41}\,{\rm erg\,s}^{-1}$ at $z{=}0.20$. Considering the range of combinations of exposure times and redshifts for our {\em Chandra} cluster survey, we find our on-axis survey limits are at or below $L_{X}{=}1.0{\times}10^{42}\,{\rm erg\,s}^{-1}$  at the cluster redshift for all 26 systems (column 6 of Table~\ref{chandra}). We expect this sensitivity to reduce by no more than 10--20\% for sources with the largest off-axis angles (${\sim}10^{\prime}$)\footnote{{\em Chandra} Proposers' Guide: {\em http://cxc.harvard.edu/proposer/}.}

The presence of extended emission from the ICM affects the ability of {\sc wavdetect} to detect faint X-ray point sources near the cluster core due to increased photon noise. This effect is demonstrated in Fig.~\ref{xrayradial} which shows the number of broadband X-ray counts as a function of cluster-centric radius for each point source matched to a galaxy counterpart with known redshift. Within 0.4\,$r_{500}$ (0.2\,$r_{500}$) {\sc wavdetect} doesn't detect any point sources with fewer than 10 (20) counts, indicating a ${\sim}$2--3${\times}$ reduction in sensitivity to X-ray point sources in the cluster cores. The impact of the ICM background on our ability to detect X-ray AGN varies significantly from cluster to cluster, being significantly worse for the cool-core clusters for which the X-ray emission is strongly concentrated in the cores (and which tend to have deeper images) than their non cool-core counterparts. We confirm that for 20/26 clusters we remain complete to at least $L_{X}{=}1.0{\times}10^{42}$\,erg\,s$^{-1}$ even in the core regions with the strongest ICM emission. We note also that all five cluster X-ray AGN identified within $0.2\,r_{500}$ have X-ray luminosities of 0.87--2.$2{\times}10^{42}$\,erg\,s$^{-1}$. For the remaining six clusters A611, A1835, A2390, RXJ\,2129.6+0005, Z2089 and Z7160 their concentrated X-ray emission related to their strong cool cores does means that we are locally (only within ${\la}0.1\,r_{500}$) insensitive to X-ray AGN at this luminosity level. 

\begin{figure}
\plotone{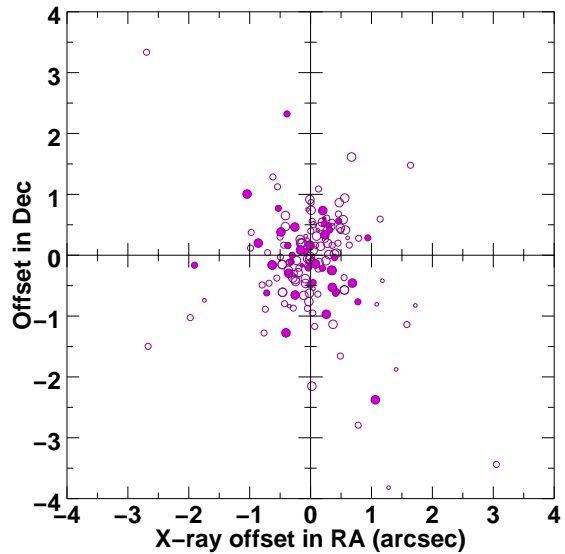}
\caption{Positional offsets ($\Delta$\,RA, $\Delta$\,Dec) between the X-ray point source and the $K$-band counterpart for all counterparts with known redshifts. Solid symbols indicate cluster members. The size of the symbol scales with the number of X-ray counts.} 
\label{xrayoffset}
\end{figure}

We identified possible NIR counterparts to the X-ray point sources, matching objects within a generous limit of 4\,arcsec. Given the positional accuracy of the X-ray centroids in {\em Chandra} images, true counterparts are expected to be within 2\,arcsec, except in the cases of large, extended galaxies for which the optical centroid may not coincide well with the active nucleus. Each possible match was then visually checked in our deep {\em Subaru} optical images, and in the case of multiple possible counterparts, the nearest match taken, unless the other is infrared-bright and within 2\,arcsec. Figure~\ref{xrayoffset} shows the positional offsets ($\Delta$\,RA, $\Delta$\,Dec) for all the optical counterparts to X-ray point sources with known redshifts. The vast majority have counterparts within 1.5\,arcsec, including 46 of 48 cluster members (shown as solid symbols), confirming that the associations are reliable.

Over the full sample of 26 clusters we have redshifts for 151 out of the 183 galaxies detected with {\em Chandra} that would be brighter than $M_{K}^{*}{+}2$ at the cluster redshift, and hence could have been targeted for spectroscopy within ACReS. This gives us a spectroscopic completeness of 83\% for the X-ray AGN subsample. We note however that only three of the remaining 32 X-ray sources would be identified as targets for spectroscopy within the ACReS selection criteria ({\S}~\ref{spectroscopy}). The rest are classified as being too red in $J{-}K$ to be cluster members (being on average ${\sim}0.2$\,mag redder than our $J{-}K$ color cut and ${\sim}0.4$\,mag redder than the cluster C-M relation), and instead are likely background galaxies. This is supported by them being on average ${\sim}1$.3\,mag fainter in the $K$-band than our spectroscopic cluster X-ray AGN sample (which has a median magnitude of $K{=}15.2$). None of these 32 X-ray sources lacking redshifts would be brighter than $L^{*}$ at the cluster redshift, whereas this is the median $K$-band luminosity of our confirmed cluster X-ray AGN sample. Thus, our effective spectroscopic completeness is ${>}95$\% for X-ray point sources hosted by galaxies brighter than $M_{K}^{*}{+}2$ whose near-IR colors are consistent with being at the cluster redshift. In terms of spatial coverage, we find that our {\em Chandra} imaging covers 90\% of cluster members within $r_{500}$ or 75\% of those within $r_{200}$ over the full sample. 

\begin{figure*}
\plotone{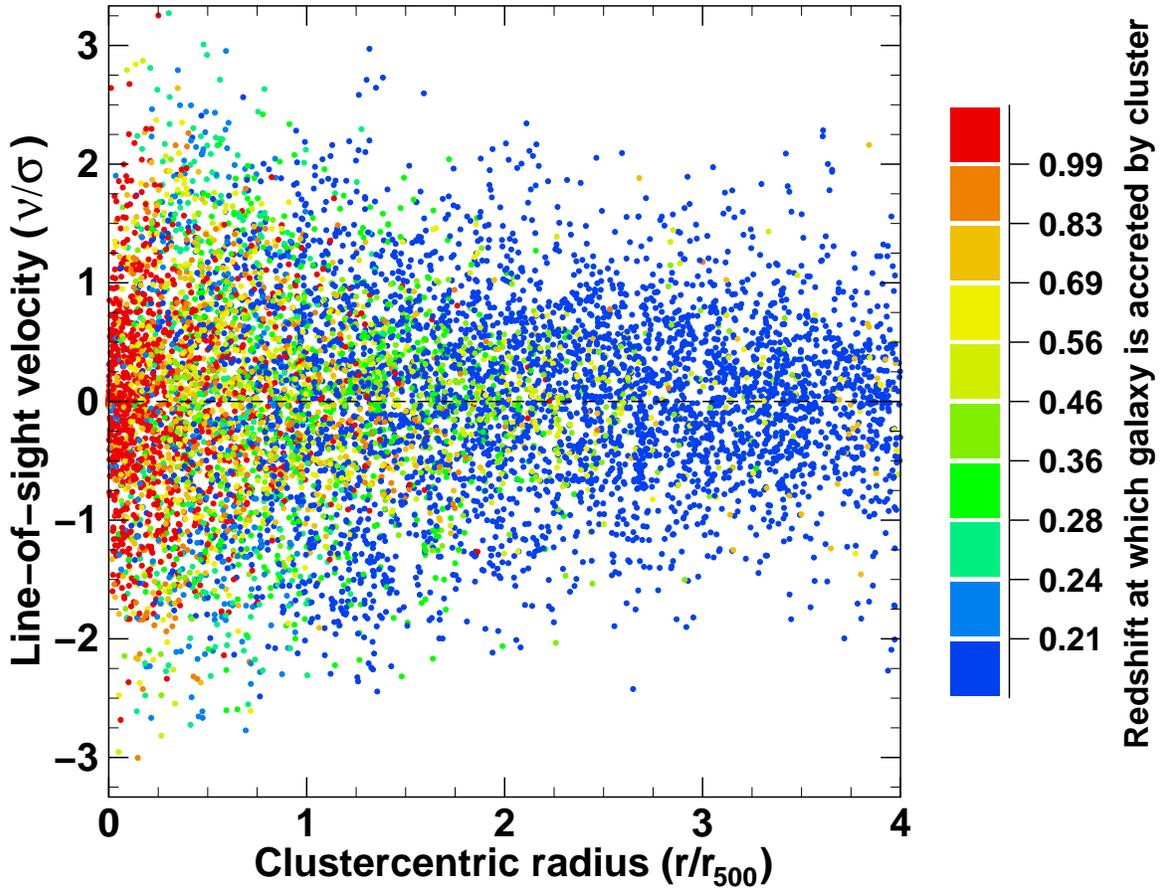}
\caption{Stacked phase-space diagram ($r_{proj}/r_{500}$ vs $(v_{los}{-}{<}v{>})/\sigma_{v}$) for the 30 most massive clusters in the Millennium simulation at $z{=}0.21$, the snapshot closest to the mean redshift of our cluster sample. Each point indicates a $M_{K}{<}{-}23.1$ galaxy from the Bower et al. (2006) semi-analytic model catalog, colored according to when it was accreted onto the cluster, those being accreted earliest indicated by red symbols. Here accretion epoch is defined as the snapshot at which the galaxy passes within $r_{500}$ for the first time. Galaxies yet to pass through $r_{500}$ by $z{=}0.21$ are indicated in mid-blue and dominate at large cluster-centric radii ($r{\ga}2\,r_{500}$) and along the caustics $|v{-}{<}v{>}|/\sigma_{v}{\ga}1$).}
\label{zspace}
\end{figure*}

We have spectroscopically identified a total of 48 member galaxies with X-ray emission in 26 clusters. For those host galaxies detected with {\em Spitzer} at 24$\mu$m, we estimated their star formation rates (SFRs) and total infrared luminosities ($L_{TIR}$) by fitting the infrared-SED models of \citet{rieke} to the observed 24$\mu$m fluxes. Table~\ref{catalog} lists their positions, redshifts and X-ray luminosities along with a summary of the key properties (M$_{K}$, SFR, morphological and spectral classes) of their host galaxies. 
 BCGs (defined here as the Brightest Cluster Galaxy in the cluster core) have been explicitly excluded due to their unique star formation histories and evolutions \citep{lin}, and the direct link between BCG activity and the presence of cooling flows within clusters \citep{edge,bildfell,smith10}. It is also unclear whether the X-ray emission comes from the galaxy itself or the ICM, particularly in the case of cooling flow clusters. 

\section{The Millennium simulation and the distribution of galaxies in the caustic diagram as a function of accretion epoch}
\label{sec:dynamic}

The location of infalling and virialized populations in and around massive clusters are well separated in radial phase-space diagrams \citep[$r$, $v_{radial}$;][]{mamon,dunner,mahajan11}. This is not the case for the observable counterpart of projected cluster-centric radius ($r_{proj}$) versus line-of-sight (LOS) velocity \mbox{($v_{los}{-}{<}v{>}$)} relative to the cluster redshift. However, via the use of cosmological simulations containing dozens of massive clusters similar to those in LoCuSS, populated by galaxies based on semi-analytic models, it is possible to produce simulated caustic diagrams for which the accretion histories of the member galaxies are known. For this purpose, $20{\times}20{\times}40\,h^{-3}{\rm Mpc}^{3}$ regions centered on the 30 most massive clusters were extracted from the Millennium Simulation \citep{springel}. These simulations cover a (500\,$h^{-1}$Mpc)$^3$ volume, producing dark matter (DM) halo and galaxy catalogs based on the semi-analytic models ({\sc galform}) of \citet{bower} for which positions, peculiar velocities, absolute magnitudes and halo masses are all provided at 63 snapshots to $z{=}0$. This allows the orbit of each galaxy with respect to the cluster to be followed from formation up until the present day, from which the epoch at which it is accreted (here defined as when it passed within $r_{500}(z)$ for the first time) can be determined. The virial masses of these 30 cluster halos at $z{=}0.21$ cover the range 2.8--21.$6{\times}10^{14}h^{-1}{\rm M}_{\odot}$, consistent with the mass range of 2.9--14.$7{\times}10^{14}h^{-1}{\rm M}_{\odot}$ for our cluster sample based on the weak lensing mass estimates of \citet{okabe09}.

Knowing the relative position and peculiar velocity of each galaxy with respect to its host cluster it is easy to reproduce observations of that cluster in the form of projected cluster-centric radii and LOS velocity. The 30 systems are then stacked, scaling each by the cluster radius ($r_{500}$) and velocity dispersion, $\sigma_{v}$. This stacking over many clusters is not required simply to match our observed sample, or even to contain sufficient galaxies to derive robust statistical properties (although this is important), but most importantly to even out all of the large line-of-sight variations due to the presence of infalling structures or filaments for individual cluster lines of sight and times of observation.  This applies both for the simulations and the observations, motivating our use of such large cluster samples. 

Figure~\ref{zspace} shows the stacked caustic plot for galaxies orbiting the 30 most massive clusters in the Millennium simulation, as would be observed at $z{=}0.21$. The x-axis shows the cluster-centric radius projected along the z-direction of the simulation, scaled by $r_{500}(z)$, while the y-axis shows the LOS velocity (along the z-axis of the simulation) of each galaxy scaled by the velocity dispersion of galaxies within that cluster. 
Each symbol indicates a $M_{K}{<}{-}23.1$ galaxy colored according to the epoch when it was accreted by the cluster. 
The overall distribution of these galaxies forms the characteristic ``trumpet''-shaped caustic profile expected for galaxies infalling and subsequently orbiting within a massive virialized structure \citep{regos,dunner}. 

The distribution of galaxies in Figure~\ref{zspace} suggests that broadly speaking, the galaxies in clusters can be thought of as belonging to two broad categories. The first population are those galaxies accreted at an early epoch, and are shown in Figure~\ref{zspace} as red symbols. These are spatially localized in the cluster core and their LOS velocities are typically ${\la}1{\sigma}$. We identify this population with those galaxies that either formed locally or were accreted when the cluster's core was being assembled, with their low velocities reflecting the fact that the system they fell into was only a fraction of the present-day mass of the cluster and possibly further slowed by dynamical friction. The second population of galaxies are those accreted after the formation of the core. Some of these galaxies are falling in for the very first time. These galaxies span the gamut from those that have yet to cross $r_{500}$ (blue symbols), those that are close to pericenter, and through to ``backsplash galaxies'' (green symbols, primarily) that have completed their first pericenter and are now outward bound \citep{mamon,pimbblet}. The maximum extent of this population in the vertical axis of Figure~\ref{zspace} is a function of both the actual velocities of the galaxies and a geometric projection factor. At small projected radii, galaxies with the highest LOS velocities are physically located deep inside the cluster, their high velocities being the result of being accelerated all the way in.

Figure~\ref{zspace} demonstrates that we can statistically associate those galaxies lying along the caustics as mostly infalling (blue symbols), and those with low LOS velocities and cluster-centric radii as the most likely to be virialized (red/orange symbols). It reveals the potential of using the caustic diagram to statistically identify the primary accretion epoch(s) of observed cluster galaxy populations based upon their distribution in the plot.

\section{Results}
\label{sec:results}

In total we have identified 48 X-ray point sources among member galaxies in the 26 clusters studied, corresponding to between 0 and 6 X-ray sources per cluster (Table~\ref{chandra}). Of these, 24 have both $L_{X}{>}10^{42}$erg\,s$^{-1}$ and $M_{K}{<}{-}23.1$ ($K^{*}{+}1.5$). From our spectroscopic survey of other cluster members within the {\em Chandra} images, we identify in total 2702 $M_{K}{<}{-}23.1$ galaxies, giving us an estimate of fraction of massive cluster galaxies ($M_{K}{<}{-}23.1$) hosting X-ray AGN ($L_{X}{>}10^{42}$erg\,s$^{-1}$) of $0.73{\pm}0.14$ percent, once we account for the fact that the spectroscopic completeness of the X-ray AGN subsample is higher than that for the remaining inactive cluster population. This is consistent with \citet{martini07} who found that $0.87{\pm}0.25$ percent of $M_{R}{<}{-}20$ galaxies in eight clusters host AGN with $L_{X}{>}10^{42}$erg\,s$^{-1}$. Although their survey is $R$-band selected rather than our $K$-band selection, both reach ${\sim}M^{*}{+}1.5$ and so should be directly comparable. Interestingly, our value is ${\sim}40$\% lower than the $1.19{\pm}0.11$ percent obtained by \citet{haggard} for $M_{R}{<}{-}20$ field galaxies at $0.05{<}z{<}0.31$ from their analysis of 323 archive {\em Chandra} images covered by SDSS DR5 imaging and spectroscopy.  

\begin{figure}
\plotone{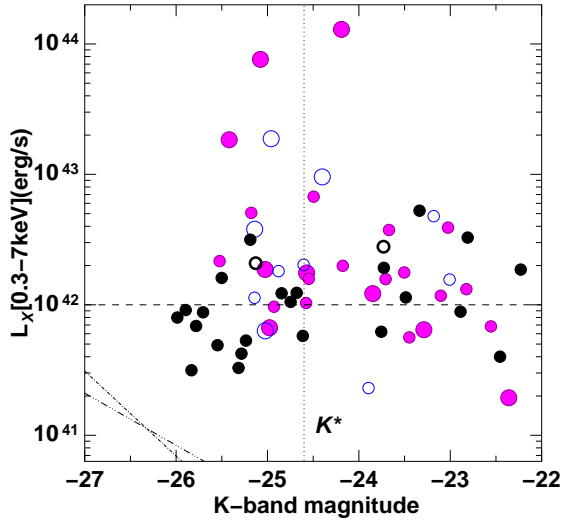}
\caption{Broadband X-ray luminosity, $L_{X}$, versus the near-infrared absolute magnitude, $M_{K}$, for X-ray detected galaxies at $0.15{<}z{<}0.30$. X-ray loud cluster members detected (not-detected) at 24$\mu$m are indicated by magenta (black) symbols, while field galaxies are shown as open blue (black) symbols. Larger symbols indicate galaxies identified as LIRGs from their 24$\mu$m fluxes. The expected levels of X-ray emission from low-mass X-ray binaries \citep[{\em dot-dot-dot-dashed line};][]{kim} or thermal emission from the diffuse gas halo \citep[{\em dot-dashed line};][]{sun} are shown.}
\label{xraylum}
\end{figure}

Thirty-two of the 48 X-ray AGN are detected with {\em Spitzer} and nine of these are identified as LIRGs (Luminous InfraRed Galaxies; $L_{TIR}{>}10^{11}L_{\odot}$, where TIR is the Total InfraRed luminosity integrated over 8--1000$\mu$m). Conversely, 6 out of the 10 most luminous cluster galaxies at $24{\mu}m$ (covered by the {\em Chandra} data) are found to be X-ray AGN. We note that this does not mean that these IR-bright AGN would be identified as AGN via the infrared selection of \citet{stern}, while the 24$\mu$m emission may be due to star formation rather than nuclear activity. We do note however that for eight of the nine LIRGs, a comparison to our available {\em Herschel} photometry reveals that they have much flatter SEDs ($(f_{100}/f_{24}){<}10$) than found for star-forming galaxies at these redshifts \citep[$(f_{100}/f_{24}){\sim}25$;][]{smith,pereira}, and more consistent with the flatter power-law SEDs of \citet{polletta} in which the infrared emission is mostly due to dust heated by nuclear activity. For the remaining 24$\mu$m-detected X-ray AGN, those objects detected by {\em Herschel} mostly have far-IR SEDs consistent with star-forming galaxies.

At X-ray luminosities $L_{X}{\la}10^{42}$erg\,s$^{-1}$ there are four potential sources of X-ray emission in galaxies: low-mass X-ray binaries (LMXBs) which are sensitive to the total stellar mass of the galaxy \citep{kim}; high-mass X-ray binaries (HMXBs) which are sensitive to the recent star formation activity \citep{mineo}; thermal emission from a hot gas halo; and an AGN. It it thus important at this point to quantify the possible contribution to the X-ray emission of our {\em Chandra} point sources from the former three components. \citet{mineo} show that HMXBs are a good tracer of recent star formation, and their collective luminosity scale linearly with the SFR, $L_{X}$(0.5--8.0\,keV$){\sim}2.5{\times}10^{39}\,{\rm SFR} ({\rm M}_{\odot}{\rm yr}^{-1})$ for SFRs in the range 0.1--1000\,M$_{\odot}{\rm yr}^{-1}$. Based on comparison of the 24$\mu$m-based SFR estimates to the observed X-ray luminosities for each point source, HMXBs contribute a maximum of 10--20\% of the X-ray luminosity, and in most cases less than 1\%. To examine the possible contribution to the X-ray emission from LMXBs or the hot gaseous coronae, which should both scale with stellar mass, Figure~\ref{xraylum} shows the broad-band X-ray luminosity against the $K$-band absolute magnitude of the X-ray sources associated to cluster members, colored according to whether they are detected at 24$\mu$m (magenta) or not (black) points. X-ray point sources associated to field galaxies at $0.15{<}z{<}0.30$ are shown in blue. The predicted relations between $L_{X}$ and M$_{K}$ for LMXBs \citep[dot-dot-dot-dashed line;][]{kim} and thermal emission from the diffuse gas halo \citep[dot-dashed line;][]{sun} are both shown, and both lie considerably below any of the X-ray point sources. From these analyses it seems likely that all of our X-ray point sources are primarily powered by central AGN rather than any other source. 

\begin{figure}
\plotone{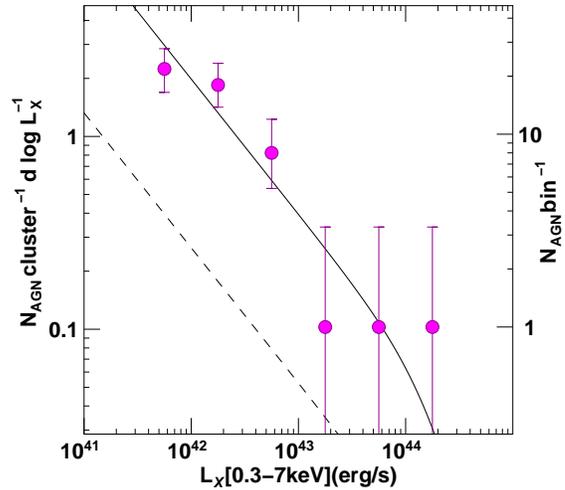}
\caption{The broadband (0.3--7\,keV) X-ray luminosity function of X-ray AGN belonging to the 26 clusters. The right-hand axis shows numbers of X-ray AGN observed per 0.5\,dex bin in luminosity, while the left-hand axis shows the averaged number of X-ray AGN per cluster within $r_{200}$ (including a 25\% correction for those galaxies within $r_{200}$ not covered by {\em Chandra} data) per dex in $L_{X}$. 
The dashed line shows the $z{=}0$.3 field X-ray luminosity function of Hasinger et al. (2005) matched to have the same volume as our cluster sample ($3.1{\times}10^{4}$Mpc$^{3}$). It has been evolved in space-density to $z{=}0.225$ and luminosity-corrected from 0.5--2\,keV to 0.3--7\,keV assuming a $\Gamma{=}1.7$ power-law spectrum. The solid curve shows the same luminosity function multiplied by a factor 7.5 to match the cluster LF.}
\label{lf}
\end{figure}

\begin{figure*}
\plotone{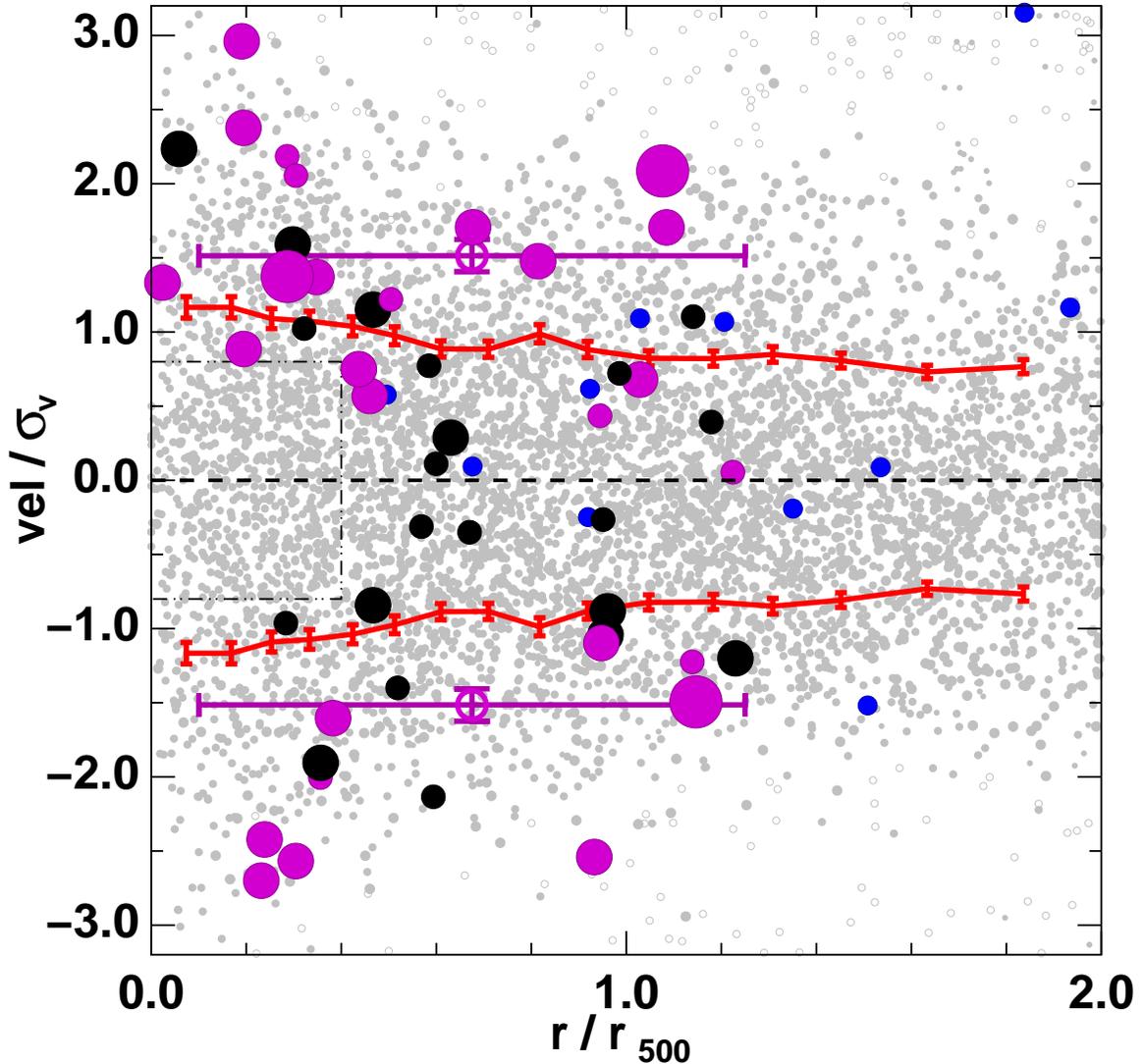}
\caption{Stacked observed phase-space diagram ($r_{proj}/r_{500}$ vs $(v_{los}{-}{<}v{>})/\sigma_{v}$) of galaxies combining all 26 clusters in our sample. Each gray solid point indicates a spectroscopic cluster member, while open points indicate field galaxies. The red curves indicate the $1{-}{\sigma}$ velocity dispersion profile in radial bins containing 200 galaxies. X-ray AGN are indicated by larger magenta (black) symbols according to whether they are detected (not-detected) with {\em Spitzer} at 24$\mu$m. The size of the magenta/black symbols correspond to X-ray luminosities, $L_{X}{<}10^{42}$\,erg\,s$^{-1}$ (smaller), $10^{42}{<}L_{X}{<}10^{43}$\,erg\,s$^{-1}$ (mid-sized), and $L_{X}{>}10^{43}$\,erg\,s$^{-1}$ (largest).
Blue symbols indicate sources spectroscopically classified as broad-line QSOs. The open magenta symbols and error bars indicate the velocity dispersion of the cluster X-ray AGN. The dot-dot-dashed box indicates the region with $r{<}0.4\,r_{500}$ and \mbox{$|v_{los}{-}{<}v{>}|/{\sigma_{v}}{<}0.80$} devoid of X-ray AGN.}
\label{caustic}
\end{figure*}

Examining Fig.~\ref{xraylum} it seems that the most X-ray luminous AGN are infrared-bright, while many of the lower luminosity X-ray AGN are massive galaxies not detected at 24$\mu$m. A Wilcoxon-Mann-Whitney U-test finds that those 24 X-ray AGN with ${\rm SFRs}{>}2{\rm M}_{\odot}{\rm yr}^{-1}$ have systematically higher $L_{X}$ than those with lower SFRs (including 24$\mu$m non-detections) at the 2.2-$\sigma$ level ($P_{U}{=}0.014$).

We find no evidence for evolution within the LoCuSS sample. Splitting the cluster sample into two according to redshift, and considering only X-ray AGN with $L_{X}{>}10^{42}$\,erg\,s$^{-1}$ where we should be reasonably complete (barring the cluster cores), we find 14 X-ray AGN in the 13 $z{<}0.227$ clusters ($f_{AGN}{=}0.78_{-0.21}^{+0.27}$\%) and 14 X-ray AGN in the 13 $z{>}0.227$ clusters ($f_{AGN}{=}0.89_{-0.24}^{+0.31}$\%). Here the AGN fractions are derived for the $M_{K}{<}{-}22.6$ cluster galaxies covered by our Chandra data.
We do find marginal evidence for a correlation with the (thermo)dynamical state of the cluster, as described by $\alpha$, the slope of the logarithmic gas density profile at 0.04\,$r_{500}$, or the cool-core strength \citep{sanderson09,smith10}. Splitting the cluster sample into two according to $\alpha$, we find 20 X-ray AGN in the 14 cool core clusters with $\alpha{<}-0.5$ ($f_{AGN}{=}1.13_{-0.25}^{+0.31}$\%), and just 8 X-ray AGN in the 12 non-cool core clusters ($\alpha{>}-0.5$ ($f_{AGN}{=}0.51_{-0.18}^{+0.25}$\%). Thus X-ray AGN are more frequent in clusters with strong cool cores at the $2{\sigma}$ level than those with weak or no cool cores. This higher frequency of X-ray AGN in cool core clusters is predominately due to those with little or no star formation (SFRs${<}2{\rm M}_{\odot}{\rm yr}^{-1}$): 9 such X-ray AGN are found in the 14 cool core clusters ($f_{AGN,\ low\ SFR}{=}0.51_{-0.17}^{+0.23}$\%) but just 2 in the 12 non-cool core clusters ($f_{AGN,\ low\ SFR}{=}0.11_{-0.07}^{+0.15}$\%). 

\subsection{The X-ray luminosity function of cluster AGN}
\label{section:lf}

Figure~\ref{lf} shows the broadband (0.3--7\,keV) X-ray luminosity function (LF) of X-ray AGN belonging to the 26 clusters. 
This confirms that ${\sim}$1--2 AGN with $L_{X}{>}10^{42}$\,erg\,s$^{-1}$ are found per cluster, consistent with \citet{gilmour}. Of these, 27/28 are moderate-luminosity AGN below $L_{X}^{*}{\sim}10^{44}$\,erg\,s$^{-1}$ \citep{hasinger}. We find just one X-ray AGN with quasar-like luminosities ($L_{X}{>}10^{44}$\,erg\,s$^{-1}$) in our sample, implying that they are extremely rare within low redshift clusters \citep[${<}0.13$ per cluster;][]{gehrels}. 
Note that none of the 32 X-ray sources lacking redshifts would have $L_{X}{>}2{\times}10^{43}$\,erg\,s$^{-1}$ if placed at the respective cluster redshifts -- i.e. the brightest two luminosity bins are robust to spectroscopic incompleteness. As discussed in {\S}~\ref{dataxray} not all of the clusters have {\em Chandra} imaging sensitive to X-ray AGN with $L_{X}{=}10^{41.5}$\,erg\,s$^{-1}$. We account for this by weighting each X-ray AGN in the lowest-luminosity bin ($41.5{<}\log L_{X}($erg\,s$^{-1}){<}42.0$) by the number of clusters in which it could be detected.

The luminosity function of X-ray AGN in clusters has the same shape as, and a normalization a factor ${\sim}7.5{\times}$ higher than, that of field X-ray AGN (dashed curve in Fig.~\ref{lf}). 
The field luminosity function is adapted from \citet{hasinger}. First we multiplied their luminosity function by the volume probed by our survey ($3.1{\times}10^{4}$\,Mpc$^{3}$). Note that for a typical cluster in our survey, the depth associated with the redshift range over which cluster membership, and thus cluster volume, is defined (e.g. $z{=}0$.195--0.222 for A\,209) is of order ${\delta}{z}{\sim}0.02$ or ${\sim}90$\,Mpc, i.e.\ much larger than the physical sizes of the clusters themselves ($r_{200}{\sim}1.$5--2.5\,Mpc). Second we took the analytical fit from their soft (0.5--2.0\,keV) X-ray LF for $0.20{<}z{<}0.40$ AGN, evolved it to our mean cluster redshift ($z{=}0.225$) based on their best-fit evolution in the space density of AGN, $(1{+}z)^{4.90}$, then corrected the X-ray luminosities from their soft band to our broad band assuming a $\Gamma{=}1.7$ power-law spectrum. 

The higher normalization of the cluster AGN luminosity function is unsurprising because clusters are overdense regions of the universe. To account for this, and thus achieve a more direct comparison between cluster and field AGN populations, we use the data from our spectroscopic survey to estimate that the redshift-space density of cluster galaxies brighter than $M_{K}{=}{-23.1}$ is a factor of ${\sim}23{\times}$ higher than that of field galaxies, when measured over the regions observed with {\em Chandra}. The net normalization of the cluster AGN luminosity function after accounting for the overdense nature of the clusters is therefore a factor of ${\sim}3$ lower than the field AGN luminosity function.

\subsection{Dynamical analysis of the X-ray AGN}
\label{section:dyn}

In Figure~\ref{caustic} we show the {\em observed} stacked caustic diagram of all 26 clusters, in which the cluster-centric radii are normalized by the $r_{500}$ determined from the {\em Chandra} data, and the LOS velocities are scaled by the velocity dispersion ($\sigma_{v}$) of all cluster members within $r_{200}$ \citep[taken to be $1.5\,r_{500}$;][]{sanderson03}.
The larger symbols indicate the X-ray AGN colored magenta (black) according to whether they are (not) detected in the mid-infrared with {\em Spitzer}, while the size of the symbols indicate the X-ray luminosity. The small gray points indicate the remaining cluster galaxies, forming the characteristic trumpet-shaped caustic profile. 

It is immediately apparent that the X-ray AGN do not trace the same distribution in the caustic diagram as the remainder of the cluster galaxy population, but preferentially trace the caustics, suggestive of an infalling population ({\S}~\ref{sec:dynamic}).
The 24$\mic$-detected AGN trace the caustics better than the non-24$\mic$ detections (we will return to this later). 

The X-ray AGN also notably avoid the area with low cluster-centric radii and relative LOS velocities, although we note that the X-ray emission from the ICM affects our ability to detect X-ray AGN at these low radii. Other spectroscopically-classified broad line QSOs (blue symbols) also appear to avoid these central regions. 
There are no X-ray AGN with $r{<}0.4\,r_{500}$ and \mbox{$|v_{los}{-}{<}v{>}|/{\sigma_{v}}{<}0.80$} (within the dot-dot-dashed box in Fig.~\ref{caustic}). In this same region of phase-space we have 423 cluster galaxies with $M_{K}{<}{-}23.1$, of which we may expect $6.1{\pm}2.5$ to be X-ray AGN by chance, based on the overall detection rate of X-ray sources (i.e. not just those with $L_{X}{>}10^{42}$\,erg\,s$^{-1}$) among the general cluster population (with {\em Chandra} coverage). The probability of detecting none is just 0.002. This region of phase-space should be dominated by those galaxies accreted earliest into the clusters ({\S}\ref{sec:dynamic}).

We do note a couple of biases which may contribute to this apparent void. Firstly, in the outer regions ($r{>}0.4\,r_{500}$) the luminosities of X-ray AGN with \mbox{$|v{-}{<}v{>}|/{\sigma_{v}}{<}0.80$} are ${\sim}3{\times}$ lower than their counterparts with higher velocity offsets. 
Secondly, the ability to detect faint X-ray AGN using {\sc wavdetect} is reduced in the cluster cores due to increased photon noise from the ICM (Fig.~\ref{xrayradial}). Hence if this trend for fainter X-ray sources to lie predominately at lower velocity offsets seen at $r{>}0.4\,r_{500}$ were to hold also in the core region, we may well be systematically missing them. However, there should be no {\em observational} bias with respect to LOS velocity. Indeed there are many X-ray AGN detected near the cluster cores ($r{\la}0.4\,r_{500}$), but all have large LOS velocities (${\sim}1$\,000--3\,700\,km\,s$^{-1}$) with respect to that of the cluster. Moreover, the fact that the optically-selected AGN also avoid the cluster core (blue symbols in Fig.~\ref{caustic}), yet shouldn't be affected by the same radial bias as the X-ray AGN, supports the view that this void is real.

\begin{figure}
\plotone{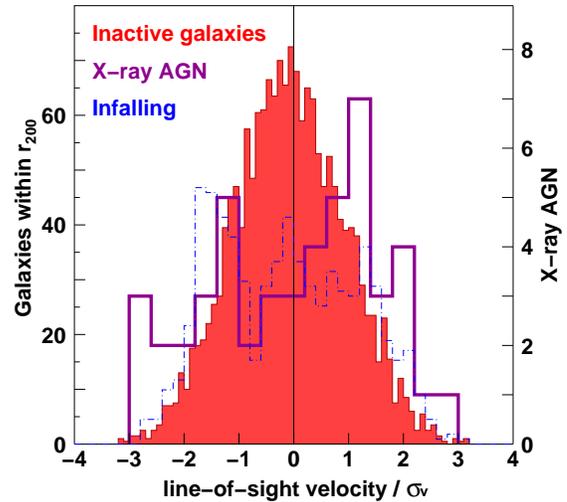}
\caption{The stacked velocity distributions (normalized by $\sigma_{v}$) of X-ray AGN (thick magenta curve) and inactive cluster members (red solid histogram) within $r_{200}$. The blue dot-dashed histogram shows the stacked velocity distribution of simulated galaxies on their first infall into the 30 most massive clusters from the Millennium simulation that lie within a projected radius of $r_{200}$ and a physical (3D) cluster-centric distance of 2\,$r_{200}$.}
\label{velhist}
\end{figure}

The mean velocity dispersion of the X-ray AGN is $1.51{\pm}0.11$ $\sigma_{v}$ (shown by open magenta symbol with error bars), significantly higher than that observed for the general cluster population (red curve). Using the Wilcoxon-Mann-Whitney non-parametric U test, we find that the absolute LOS normalized velocities \mbox{$|v_{los}{-}{<}v{>}|/\sigma_{v}$} of the 48 X-ray AGN are systematically higher than the remainder of the cluster population (just those covered by our {\em Chandra} images) at the 4.66-$\sigma$ significance level ($P_{U}{\sim}1.6{\times}10^{-6}$). In contrast we find no statistical difference in the radial distribution of the X-ray AGN from the remainder of the cluster population. Comparing visually the X-ray AGN detected with {\em Spitzer} (magenta points) and those not detected (black points), there is a suggestion that the {\em Spitzer} detections have systematically higher absolute line-of-sight velocities. A Wilcoxon-Mann-Whitney test finds those 23 X-ray AGN with obscured SFRs${>}$2\,M$_{\odot}$\,yr$^{-1}$ (or $L_{TIR}{>}2{\times}10^{10}L_{\odot}$) have systematically higher \mbox{$|v{-}{<}v{>}|/\sigma_{v}$} than those with lower SFRs at the $2.3{\sigma}$ level. Similarly, the more X-ray luminous AGN (indicated by the larger symbols) appear to have higher velocity offsets than their lower luminosity counterparts. Again, a Wilcoxon-Mann-Whitney test finds that the 29 X-ray AGN with $L_{X}{>}10^{42}$\,erg\,s$^{-1}$ have systematically higher \mbox{$|v{-}{<}v{>}|/\sigma_{v}$} than those with lower X-ray luminosities at the $2.5{\sigma}$ level ($P_{U}{=}0.005$).

\begin{figure*}
\plotone{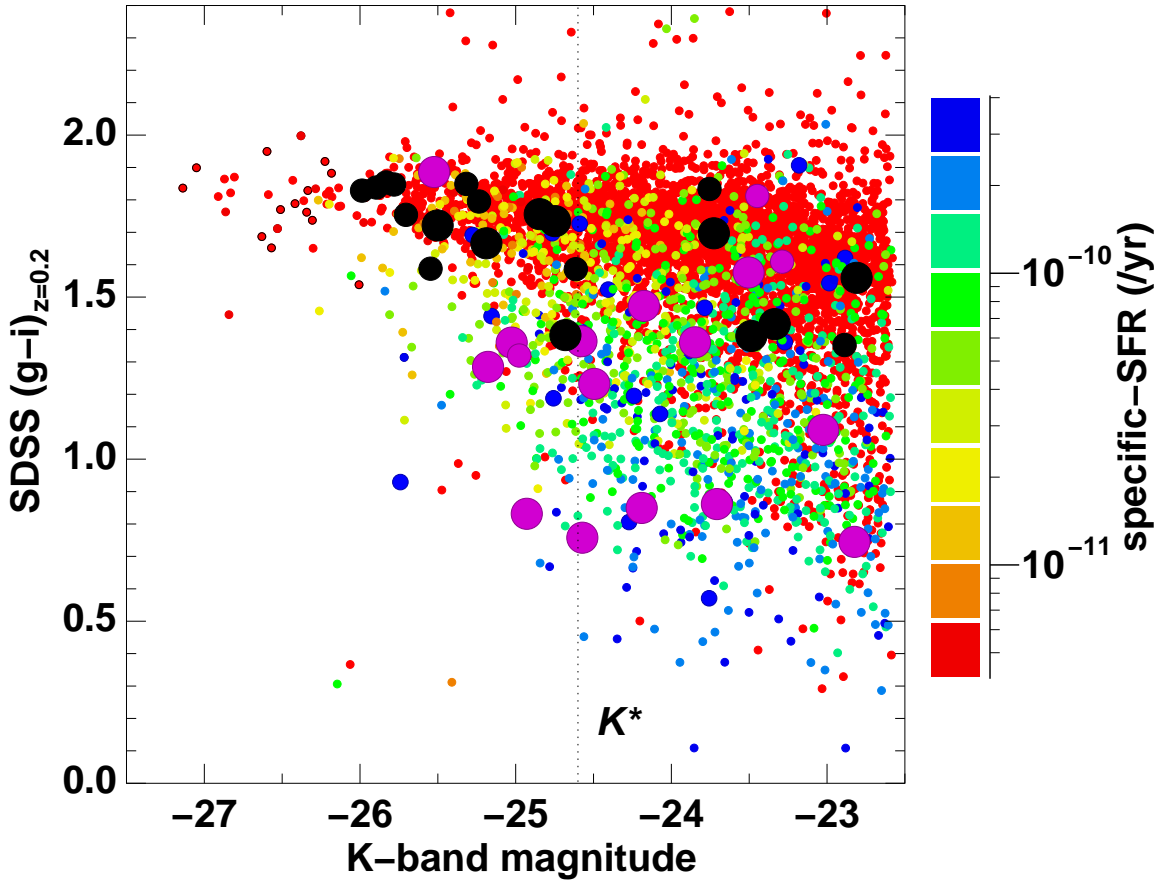}
\caption{Color-magnitude ($g{-}i/M_{K}$) diagram of cluster galaxies (small symbols), color coded according to their specific-SFRs (combining the 24$\mu$m-based SFRs estimated using the calibration of \citet{rieke} with stellar masses derived from M$_{K}$). The $g{-}i$ color of each cluster galaxy is k-corrected to $z{=}0.2$ enabling all of the clusters to be stacked onto a single combination of a red sequence of passive galaxies not detected at 24$\mu$m (red points) and blue cloud of star-forming galaxies (green/blue points). X-ray AGN are indicated by larger magenta (black) symbols according to whether they are detected (not-detected) with {\em Spitzer} at 24$\mu$m. Large blue symbols indicate sources spectroscopically identified as broad-line QSOs. Seven X-ray AGN do not have SDSS coverage and three more are fainter than M$_{K}^{*}{+}2$ and are not plotted.}
\label{cmdiagram}
\end{figure*}

The velocity distributions of X-ray AGN and inactive (not X-ray point sources or 24$\mu$m-detected) cluster galaxies for our stacked cluster sample are compared directly in Figure~\ref{velhist}. While the velocity distribution of inactive galaxies within $r_{200}$ (solid red histogram) can be described approximately as a Gaussian function of width $\sigma_{v}$, the velocity distribution of X-ray AGN (magenta curve) appears more consistent with a flat, top-hat profile than a Gaussian. The kurtosis of the velocity distribution of X-ray AGN (1.965) is significantly lower than that expected for a Gaussian distribution (3.0) at the 98.9\% confidence level, but close to expectations from a uniform distribution (1.8). 

We may be concerned that many of the X-ray AGN at high absolute LOS velocities are not in fact cluster galaxies, but interlopers from the general field population. However, we identify just 14 X-ray AGN in the general field population at $0.15{<}z{<}0.30$ in the same {\em Chandra} images, that is after excluding the redshift ranges deemed to contain the cluster populations. Based on this field X-ray AGN space density, we would expect ${\sim}3$ field X-ray AGN to be interlopers within the cluster population by chance for the whole survey of 26 clusters, giving an estimated field contamination of 6\%. Hence it seems unlikely that the bulk of the X-ray AGN located along the caustics in Fig.~\ref{caustic} are interlopers and there by chance due to projection effects, but are instead a genuine population of galaxies infalling into the clusters for the first time. The highest velocities of cluster galaxies occur for those galaxies on their first cluster infall, having passed through the virial radius on predominately radial orbits. 

From this comparison to the field population, and the observed X-ray luminosity function ({\S}~\ref{section:lf}), the density of X-ray AGN in redshift-space within clusters appears ${\sim}$8--1$6{\times}$ higher than that for the general field. This does not mean that the encountering of the cluster environment is somehow ``triggering'' the X-ray AGN, but simply reflects the much higher space densities of galaxies as a whole within clusters. Indeed, the fraction of cluster members with $M_{K}{<}{-}23.1$ identified as an X-ray point source (of any $L_{X}$) is $1.4{\pm}0.2$\%, is marginally lower than that observed for the $0.15{<}z{<}0.30$ field galaxy population in the same {\em Chandra} images ($1.8{\pm}0.5$\%). 

\begin{figure*}
\plotone{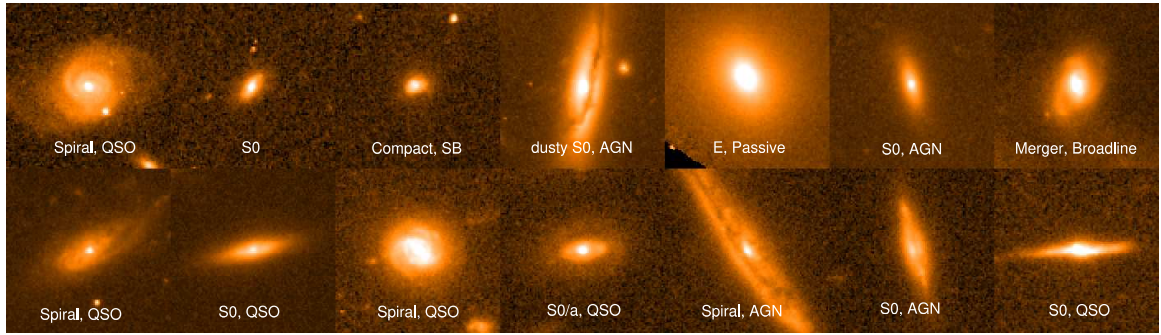}
\caption{Thumbnail {\em HST} images of all 14 X-ray AGN from our cluster sample for which {\em HST} imaging is available. The visual morphology and spectral classification are indicated for each host galaxy where available.}
\label{mosaic}
\end{figure*}

\subsection{Optical colors, spectra and morphologies}

We have spectral coverage for 42 of the 48 X-ray AGN from our ACReS MMT/Hectospec dataset, the remaining six redshifts coming from the literature. Of those for which we have spectra: 16 ($38.1^{+8.9}_{-8.2}$\%) show broad-line H$\alpha$ emission of Type I Seyferts/QSOs; 17 more ($40.5^{+8.9}_{-8.4}$\%) show narrow-line H$\alpha$ emission which may be due to either nuclear activity or star formation; while 9 ($21.4^{+8.2}_{-6.7}$\%) appear passive, lacking any clear emission lines. All of the Type I AGN are detected at 24$\mu$m as opposed to just 3/9 of the passive galaxies.

Figure~\ref{cmdiagram} shows the optical color-magnitude ($g{-}i/M_{K}$) diagram of cluster galaxies for the 22 clusters for which SDSS DR7 $ugriz$ photometry is available. 
The IR-dim X-ray AGN (black symbols) lie on or close to the cluster red sequence, suggesting little impact from the nuclear activity on galaxy color. The 24$\mu$m-detected X-ray AGN (magenta symbols) mostly lie within the blue cloud, suggesting star-formation concurrent with nuclear activity. These objects appear blue due to their continuum flux, rather than AGN emission lines affecting their colors. 

The morphological classification correlates strongly with location within or below the cluster red sequence. All but one of the 24$\mic$-detected X-ray AGN below the red sequence are spirals or unresolved, while all of the ${\ga}L^{*}$ X-ray AGN along red sequence are E/S0s. Overall, we estimate that 24/48 ($50{\pm}8.2$\%) of the X-ray AGN are early-types (E/S0s), 8 are undisturbed spirals ($16.7^{+7.2}_{-5.6}$\%), 9 disturbed spirals/mergers ($18.8^{+7.4}_{-5.8}$\%) and 7 unresolved objects ($14.5^{+7.0}_{-5.2}$\%), which are either low-luminosity galaxies or QSOs. 
To determine morphological classifications we have high-resolution optical imaging from the {\em Hubble Space Telescope} for 14 out of the 48 X-ray AGN identified as cluster members, thumbnails of which are shown in Fig.~\ref{mosaic}. Eleven of the 14 are detected with {\em Spitzer}, five of which appear to be spirals, plus one merger. The three not detected with {\em Spitzer} appear as early-types (two S0s, one E). For the remaining 34 X-ray AGN we have high-quality Subaru optical imaging which affords visual morphological classifications, albeit with difficultly in distinguishing S0s and early-type spirals as spiral arms get smoothed out, so some spirals may be mis-classified as early-types. The distribution of morphologies shown in the HST thumbnails are thus representative of the whole sample. 

All of the undisturbed and disturbed spirals and mergers are detected at 24$\mu$m, as are 5/7 of the unresolved objects. Of the 32 X-ray AGN detected at 24$\mu$m, 17 are spirals (both undisturbed/disturbed), five are unresolved and the remaining 10 are early-type galaxies, of which all five with HST imaging were classified as S0s. We note that four of these 10 early-types show broad-line optical emission. Of those 16 X-ray AGN not detected with {\em Spitzer}, 14 are morphologically identified with early-types, the other two being unresolved. 

\begin{figure*}
\plotone{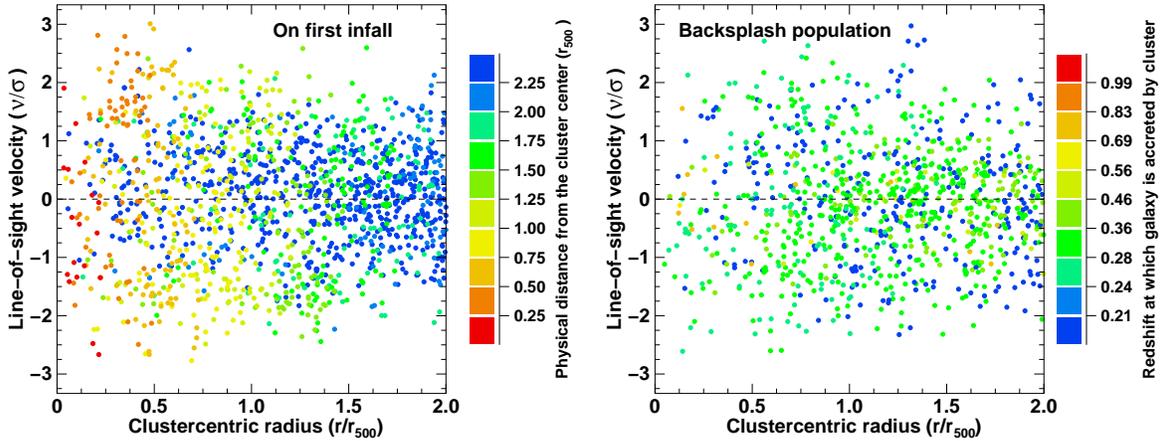}
\caption{Comparison of the locations in the phase-space diagram ($r_{proj}/r_{500}$ vs \mbox{$(v_{los}{-}{<}v{>})/\sigma_{v}$}) of infalling (left panel) and back-splash (right panel) cluster galaxies from the stacked cluster sample from the Millennium simulation. Infalling galaxies are defined as those with negative radial velocities which have yet to reach pericenter in their orbit around the cluster. They are color coded according to their physical (3D) distance from the cluster center from red ($r{<}0.25\,r_{500}$) to blue ($r{>}2.0\,r_{500}$).  
Back-splash galaxies are defined as those which have passed through the pericenter for the first time, and their orbits are now taking them back out from the cluster towards apocenter. Their symbols are instead color coded as in Fig.~\ref{zspace}.}
\label{backsplash}
\end{figure*}

\section{Discussion}
\label{sec:discuss}

We have studied the distribution of X-ray AGN in 26 massive clusters at $0.15{<}z{<}0.30$ from LoCuSS, combining {\em Chandra} imaging sensitive to X-ray sources of luminosity $L_{X}{\sim}10^{42}$\,erg\,s$^{-1}$ at the cluster redshift, with extensive and highly complete spectroscopy of cluster members. In total 48 X-ray AGN were identified among the cluster members, with luminosities $2{\times}10^{41}-1{\times}10^{44}$\,erg\,s$^{-1}$. 

\subsection{The dynamical status of X-ray AGN within clusters}

The principal result of our analysis of the X-ray AGN population within the 26 clusters was shown in Fig.~\ref{caustic}, which showed their location within the stacked caustic diagram, $(v_{los}{-}{<}v{>})/\sigma_{v}$ versus $r_{proj}/r_{500}$. They are preferentially located along the caustics, suggestive of an infalling population. They also appear to avoid the region with the lowest radii and relative velocities ($r{<}0.4\,r_{500}$; $|v_{los}{-}{<}v{>}|/{\sigma_{v}}{<}0.80$; dot-dot-dashed box in Fig.~\ref{caustic}), which is dominated by the virialized population of galaxies accreted earliest by the clusters. This dynamical behavior also shown by the optically-selected Type I Seyferts/QSOs, which also make up $38.1_{-8.2}^{+8.9}$\% of the X-ray AGN sample. Based on the simulated stacked caustic plots of the 30 most massive clusters in the Millennium simulations at $z{=}0.21$ (Fig.~\ref{zspace}), 86 (71) percent of all galaxies with $r{<}0.4\,r_{500}$ and $|v_{los}{-}{<}v{>}|/{\sigma_{v}}{<}0.80$ were accreted more than 1.5 (3.7)\,Gyr previously, while just 5\% are infalling galaxies yet to pass within $r_{500}$.

The velocity dispersion of the X-ray AGN population is ${\sim}5$0\% higher than the overall cluster population. This kinematic segregation is significant at the 4.66-$\sigma$ level. Considering a simple kinematical treatment of infalling and virialized cluster galaxies in a cluster-scale potential well leads to $|T/V|{\approx}1$ for infalling galaxies and $|T/V|{\approx}1/2$ for the virialized population, where T and V are the kinetic and potential energies. Thus, the velocity dispersions of infalling and virialized galaxies are related by $\sigma_{infall}{\approx}\sqrt{2}\,\sigma_{virial}$ \citep{colless}, close to the ratio observed between the X-ray AGN and inactive populations. For our simulated cluster galaxies, the highest velocity dispersions are seen for galaxies which have been accreted by the cluster within the last 700\,Myr, passing through $r_{500}$ for the first time between $0.21{<}z{<}0.28$ (light blue/turquoise points in Fig.~\ref{zspace}), reaching ${\sim}1.5{\sigma}_{v}$, i.e. similar to that seen for our X-ray AGN. The velocity dispersion rapidly drops back to ${\la}1.0{\sigma}_{v}$ for galaxies accreted earlier into the cluster, and is also too low (${\sim}1.0{\sigma}_{v}$) for those yet to pass within $r_{500}$ for the first time. 

The velocity histogram (Fig.~\ref{velhist}) shows a relatively flat distribution, unlike the approximately Gaussian profile of the inactive galaxy population, as confirmed by its measured kurtosis being lower than that expected for a Gaussian distribution at the 98.9\% confidence level. \citet{sanchis} show that high velocity dispersions and low kurtosis values are predictors of galaxies on predominately radial orbits, at least in the cluster cores. This flat-topped distribution is also seen for galaxies within a projected radius of $r_{200}$ that are infalling into our simulated clusters for the first time (blue dot-dashed histogram in Fig.~\ref{velhist}), and yet to reach the pericenter of their orbit. 

Both these velocity profiles and the kinematic segregation of the X-ray AGN in the stacked caustic profiles are similar to those observed for spiral and/or emission-line galaxies within clusters \citep[e.g.][]{biviano97,biviano02,boselli,haines10}. From the first dynamical studies of cluster galaxies, the velocity dispersions of spiral galaxies have been found to be systematically higher than early-types \citep{tammann, moss}. Based on much larger cluster samples, \citet{biviano97}, \citet{adami} and \citet{aguerri} all found that the stacked velocity dispersions of blue/emission-line galaxies to be on average 20\% higher than the remaining inactive galaxies.  \citet{biviano97} also examined the kinematics of spectroscopic AGN in local clusters, finding them to have systematically higher velocity offsets than the inactive galaxy population. \citet{biviano} later showed that if early-type galaxies are assumed to have isotropic orbits within clusters as supported by the Gaussian shape of their velocity distribution, the kinematic properties of late-type spirals are inconsistent with being isotropic at the ${>}9$9\% level. Instead they indicate that spirals and emission-line galaxies follow radial orbits in clusters, pointing towards many of them being on their first cluster infall. 

The cluster environment is known to strongly impact the evolution of member galaxies, transforming infalling star-forming spiral galaxies into passive early-type galaxies, as manifest by the SF--density \citep{dressler} and morphology--density \citep{dressler80} relations. The kinematic segregation of star-forming and passive galaxies in clusters is a key empirical foundation for our understanding of this process, by revealing that, while the passive galaxies are a virialized population that have resided within the cluster for many Gyr, the late-type star-forming galaxies are much more recent arrivals, and indeed many have yet to encounter the dense intra-cluster medium. It is believed that this arrival of the spiral galaxy within the cluster and its passage through the ICM removes its gas supply, leading to the subsequent quenching of its star formation \citep[and references therein]{boselli}. That the kinematic signature of the X-ray AGN within clusters is so similar to that of star-forming spirals, indicates that they are also recent arrivals into the cluster. Their apparent absence in the central regions dominated by those galaxies accreted earliest into clusters, confirms that the harsh cluster environment strongly suppresses radiatively-efficient nuclear activity. 

We find further dynamical support for the ongoing suppression of nuclear activity in galaxies as they arrive in the clusters, with both X-ray bright ($L_{X}{>}10^{42}$\,erg\,s$^{-1}$) and IR-bright ($L_{TIR}{>}2{\times}10^{10}L_{\odot}$) sub-samples of X-ray AGN showing higher velocity dispersions than their X-ray dim and IR-dim counterparts at the ${>}2{\sigma}$ level. This is consistent with the nuclear activity responsible for the X-ray and infrared emission being gradually shut down as the host galaxies are accreted into the cluster, losing their orbital kinetic energy as part of the long dynamical process of becoming virialized cluster members. We note however, that the typical duty cycle of AGN of $10^{7}{-}10^{8}$\,yr is much lower than the ${\ga}$1--2${\times}10^{9}$\,yr cluster crossing time-scales which are likely required for such velocity segregation to occur \citep[e.g.][]{gill}. Instead we suggest that the velocity segregation reflects the slow decline in the gas contents of the host galaxies available for intermittently fuelling nuclear activity.

We attempt to confirm this viewpoint by going back to the stacked caustic plot for galaxies orbiting the 30 most massive clusters in the Millennium simulation discussed in {\S}~\ref{sec:dynamic}. In Figure~\ref{backsplash} we replot the caustic diagram of Figure~\ref{zspace}, but now split the galaxies into specific sub-populations based on their orbital paths. 

In the left-hand panel we show only those galaxies which are on their first infall into the cluster, that is they have negative radial velocities and are yet to reach pericenter for the first time. Each point is color coded according to its physical (3D) distance from the cluster center in units of $r_{500}$. We see that these galaxies show a wide range of line-of-sight velocities, evenly distributed over the phase-space with a slight preference towards the caustics at ${\sim}{\pm}2{\sigma}_{v}$, as shown previously by their flat-topped velocity dispersion in Figure~\ref{velhist}. This was found to be the sub-population with the distribution of galaxies in the caustic diagram which was the most qualitatively similar to that seen for the X-ray AGN detected by {\em Spitzer} (the magenta points in Fig.~\ref{caustic}). We note that although there is no concentration of galaxies towards low relative LOS velocities and radii (as seen for the earliest accreted galaxies), they do not entirely avoid this region either (as our observed X-ray AGN apparently do). This could only be achieved by removing those galaxies within physical radii of ${\la}0$.5\,$r_{500}$ (red, orange points) and also those at ${\ga}2\,r_{500}$ (blue points). However this would also impact the required presence of galaxies at ${\pm}2{\sigma}_{v}$ and low projected cluster-centric radii (${\la}0.4\,r_{500}$). We believe that many of the X-ray AGN are near pericenter as required to achieve their high LOS velocities.  

The galaxies with the highest LOS velocities along the caustics at ${\sim}{\pm}2{\sigma}_{v}$ are found to be infalling galaxies at ${\sim}0$.25--1.0\,$r_{500}$ (orange, yellow points). Such high LOS velocities are notably not seen for those infalling galaxies at large physical distances from the cluster (${\ga}2\,r_{500}$; blue points). These galaxies have detached from the Hubble flow but haven't fallen sufficiently far into the cluster gravitational potential well to attain the very high infall velocities required to be located along the caustics, resulting in them having relative line-of-sight velocities within ${\sim}1.0{\sigma}_{v}$. This could suggest that the nuclear activity is being triggered in these infra-red bright galaxies as they are accreted into the cluster for the first time, at ${\sim}$1--2\,$r_{500}$ from the cluster center, when the galaxies have reached close to their maximal velocities, or indeed at pericenter passage. This phase could be induced by galaxy harassment whereby infalling spiral galaxies undergo frequent high-speed fly-by encounters with other cluster members, or tidal shocks as they pass through the cluster core, driving instabilities that funnel gas into the central regions triggering nuclear activity \citep{moore}. This would be consistent with our finding that all but one of the IR-bright X-ray AGN are spirals or unresolved. Unfortunately, as our {\em Chandra} data do not extend into the infall regions beyond 1--1.5\,$r_{500}$, we cannot easily test whether there are extra X-ray AGN among the infalling population just at these cluster-centric radii (as opposed to further out in the infall regions).

In the right-hand panel we show the back-splash population of galaxies which have passed through the cluster core for the first time and are on their way back out of the cluster. That is they are between first pericenter and apocenter along their orbits and have positive radial velocities. In terms of cluster-centric radius this population peaks at ${\sim}1\,r_{500}$, becoming steadily less frequent towards the lowest projected radii. This is due to them having their highest orbital velocities at pericenter, slowing down as they approach apocenter and hence spending more time near this point. Although the spread of LOS velocities for the back-splash population is similar to that for the infalling galaxies, they are more concentrated towards low LOS velocities, as seen previously by \citet{gill} and \citet{mahajan11}. We believe this distribution more closely resembles that seen for our X-ray AGN not detected with {\em Spitzer} (black points in Fig.~\ref{caustic}). This back-splash population is observed ${\sim}1$\,Gyr later after accretion into the cluster than the infalling population, as indicated by their predominately green symbols in the Figure. This suggests that as the galaxy passes through the cluster core for the first time, the gas available for star formation and its associated infrared emission is removed via ram-pressure stripping, while the gas available for nuclear activity, being deeper in the galaxy potential is harder to sweep out, allowing nuclear activity to continue past pericenter passage. The different morphologies of the galaxies hosting the IR-bright (mostly spirals) and IR-dim (mostly E/S0s) X-ray AGN suggests however that the primary factor in determining whether the host galaxy is IR-luminous is the global availability of gas and dust in the galaxy host itself, IR-dim X-ray AGN being hosted in early-type galaxies which have sufficient gas in the nucleus to fuel the AGN, but not enough on larger scales to feed obscured star formation.

The results of our dynamical analysis of X-ray AGN are in marked contrast to the comparable survey of \citep{martini07} who found that the spatial and kinematical distribution of their X-ray AGN sample were fully consistent with being drawn from the inactive cluster population. Given that their sample size is similar to ours, it is surprising that the results are so different and apparently inconsistent. The primary difference between the surveys is that their X-ray AGN are systematically less luminous than ours. Hence, in the scenario described above in which nuclear activity is slowly quenched when the host galaxy is accreted into the cluster, the low-luminosity X-ray AGN that dominate the sample of \citet{martini07} may have witnessed the cluster environment for longer than those in our sample, and so are less kinematically distinct from the general cluster population. Interestingly, \citet{ajello} find that the fraction of Seyfert 2 objects among their all-sky hard X-ray AGN sample is much higher within the largest concentrations of matter in the local Universe, suggesting that the broad-line Type I AGN are being preferentially shut down in these dense environments, consistent broadly with our kinematic segregation of IR-bright/IR-dim X-ray AGN.

We note finally that despite our result that the X-ray AGN are an infalling population (or just after pericenter) which is strongly {\em suppressed} by the cluster environment, they represent a ${\sim}$8--1$6{\times}$ overdensity in redshift space with respect to the general field, as revealed by the X-ray luminosity function of cluster AGN ({\S}\ref{section:lf}). This is fully consistent with the statistical overdensities of X-ray AGN previously observed toward clusters with respect to non-cluster fields \citep[e.g.][]{cappi,molnar,ruderman}. This does not imply that they are somehow triggered by the cluster environment, but instead reflects the overall significant (${\sim}1$0--2$00{\times}$) overdensities of galaxies in the {\em infall regions} of clusters, that have yet to encounter the ICM. Indeed the redshift space density of normal cluster galaxies in our sample is ${\sim}23{\times}$ higher than of field galaxies, higher than the respective overdensity of cluster X-ray AGN. 
It is still possible that there is an increased frequency of nuclear activity among these infalling galaxies, due to encounters between galaxies in the connecting filamentary web, or pre-processing within infalling groups, although our lack of {\em Chandra} coverage of these infall regions prevents us from measuring this. Nuclear activity may also be triggered by their interaction with the cluster itself, either as they pass through virial shocks, via compression of gas onto the nucleus in the early stages of ram-pressure stripping, or tidal shocks as they pass through cluster pericenter. 

\subsection{Comparison to results from clustering analyses}

A complementary approach to constraining the typical environment of AGN, and gain insights into the physical conditions of the accretion onto SMBHs, is to measure the spatial clustering of X-ray AGN to estimate the average DM halo mass harboring the AGN and the relative frequency of being hosted by central or satellite galaxies. 
Numerous clustering analyses have revealed a consistent view that X-ray AGN are hosted in DM halos of masses ${\sim}2{\times}10^{13}{\rm M}_{\odot}$, typical of poor groups, at all redshifts up to $z{\sim}2$ \citep{hickox,cappelluti,allevato,miyaji}. \citet{starikova} show that they are predominately located at the centers of DM halos of mass ${\ga}6{\times}10^{12}{\rm M}_{\odot}$, and tend to avoid satellite galaxies in comparable or more massive halos, fixing the limit to the fraction of AGN in satellite galaxies to be ${<}1$2\% (90\% confidence limit). Similarly, \citet{hickox} find the clustering of X-ray AGN to be consistent with that of typical galaxies on scales of 1--10\,Mpc, but significantly anti-biased on small scales (0.3--1\,Mpc), which they explain as due to their preferential location within central galaxies. \citet{tasse} also find that on 450\,kpc scales the most luminous X-ray AGN ($L_{X}{>}10^{43}$\,erg\,s$^{-1}$) are found in underdense environments in comparison to normal galaxies of the same stellar mass. 

In direct contrast, \citet{mountrichas} found X-ray AGN to be {\em more} clustered than galaxies at all scales, with no evidence for anti-bias on small scales. While they also obtained a typical host DM halo mass of ${\sim}10^{13}h^{-1}{\rm M}_{\odot}$, the stellar masses of their host galaxies were lower than expected for the typical central galaxy of such a halo, suggesting that they are associated to satellite galaxies. \citet{miyaji} found excess clustering on small scales between X-ray AGN and luminous red galaxies, consistent with a significant fraction of X-ray AGN being hosted in satellite galaxies. 

\citet{miyaji} also indicate that the AGN fraction in groups and clusters declines with halo mass. \citet{arnold} also find that the X-ray AGN fraction ($L_{X}{>}10^{41}$\,erg\,s$^{-1}$) declines by a factor two from groups ($f_{AGN}{=}0.091^{+0.049}_{-0.034}$) to clusters ($f_{AGN}{=}0.047^{+0.023}_{-0.016}$) from their study of 16 systems at $0.02{<}z{<}0.06$ . This trend remained when considering only early-type galaxies, and so was independent of any differences in the morphological mix between groups and clusters. \citet{hwang} also observed a decline in AGN fraction from groups to clusters (including at fixed morphology) for optically-selected AGN.

The observed dynamics of the X-ray AGN in our cluster sample are consistent with their being entirely drawn from an infalling population, with essentially no component coming from virialized satellite galaxies within the cluster halo. These results are consistent with the clustering analyses of \citet{hickox} and \citet{starikova}, associating the X-ray AGN with central galaxies. Their reported anti-bias on small scales then reflects our observed suppression of nuclear activity among the virialized cluster galaxy population. It is instead hard to reconcile our results with the findings of \citet{miyaji} and \citet{mountrichas} that X-ray AGN being even more clustered on small scales than the already strongly clustered luminous red galaxies, as in this case they should be preferentially located in the cluster cores rather than avoiding them as we find.

\subsection{The properties of the galaxies hosting X-ray AGN in clusters}

Studying the physical properties of the galaxies which host AGN provide complementary clues as to their accretion processes. The stellar masses, bulge masses or stellar velocity dispersions can be used to estimate the mass of the central SMBH via the known tight correlations, from which the Eddington ratio can be derived. In the absence of direct measures of the H{\sc i} or molecular gas contents, galaxy colors can be used as a proxy for star-formation history, and hence the availability of gas. Finally, galaxy morphologies provide constraints on the stage of any ongoing merging event (plus mass ratio) or the presence of bars or disks required for the related secular processes. These probes are however more prone to systematics, as the AGN can easily outshine the hosts, affecting not only their luminosities but also colors. 

\citet{hickox} found that while radio AGN are hosted mainly by massive, red sequence galaxies, the X-ray and infrared-selected AGN are instead both found in ${\sim}L^{*}$ galaxies, with the X-ray population being preferentially ``green valley'' objects, while IR AGN are slightly bluer. \citet{haggard} also found that X-ray AGN are much more likely to be located within the blue cloud or green valley than the red sequence. \citet{georgakakis09} suggest that the color distribution for X-ray AGN hasn't evolved between $z{\sim}0.8$ and the present day. The optical colors and luminosities of our cluster X-ray AGN are fully consistent with these field samples, being hosted typically by ${\sim}L^{*}$ galaxies located both within the red sequence and the blue cloud (Fig.~\ref{cmdiagram}). We find marginally more X-ray AGN along the cluster red sequence ($49{\pm}9$\%) than \citet{hickox} ($39{\pm}4$\%), which could reflect a cluster-specific difference in the host properties, such as reduced levels of star formation, but more likely reflects our use of $g{-}i$ color rather than $u{-}r$ to define the red sequence, and the systematically lower X-ray luminosities in our sample.  
The IR-dim X-ray AGN lie on or close to the cluster red sequence in Fig.~\ref{cmdiagram}, suggesting little impact from the nuclear activity on galaxy color. We find the 24$\mu$m-detected X-ray AGN to lie mostly within the blue cloud, consistent with \citet{hickox}, and suggesting star-formation concurrent with nuclear activity.

From our spectroscopic analysis of the cluster X-ray AGN, we found that 38\% show broad-line H$\alpha$ emission of Type I Seyferts/QSOs, 41\% show narrow-line H$\alpha$ emission which may be due to either nuclear activity of star formation, while 21\% appear passive, lacking any clear emission lines. These fractions are strongly inconsistent with those obtained by \citet{martini} from their spectroscopic survey of X-ray AGN in clusters. They found that only 4 out of 40 X-ray AGN showed clear optical signatures of nuclear activity in the form of significant O{\sc ii}, O{\sc iii} or broad H$\alpha$ emission. The remaining $90^{+5}_{-7}$\% show only modest star formation or appear to be passively evolving galaxies. The primary difference between the two samples is that the X-ray AGN from Martini et al.\ are systematically ${\sim}2.5{\times}$ less luminous than ours, having a mean value of $\log(L_{X}){=}41.78$ as opposed to our mean value of $\log(L_{X}){=}42.16$. We would thus expect the levels of optical emission to be correspondingly lower from the unified AGN model. \citet{burlon} find that low-luminosity X-ray AGN are more likely to be absorbed ($N_{H}{>}10^{22}$\,cm$^{-2}$), which may further contribute to the lack of optical emission among the X-ray AGN sample of \citet{martini}.

\citet{cisternas} found that $<$15\% of X-ray AGN in the HST-COSMOS field (with $L_{X}{\sim}10^{43.5}$\,erg\,s$^{-1}$; $z{\sim}0$.3--1.0) showed any signs of distortions indicative of recent mergers, and indeed found no statistical difference in the distortion fractions between X-ray AGN and inactive galaxies, indicating that major mergers are not the most relevant mechanism for the triggering of X-ray AGN at $z{\la}1$. Instead they found that over 55\% of the X-ray AGN are hosted by disk galaxies, and suggest that the bulk of black hole accretion occurs through internal secular fuelling processes and minor mergers. Similarly, \citet{griffith} found while radio-loud AGN are mostly hosted by early-type galaxies, X-ray AGN are mostly either disk-dominated (31--46\%) or unresolved point sources (31--61\%) with few (9--21\%) hosted by bulge-dominated systems. 
In the local Universe, \citet{koss11} found that X-ray AGN are ${\sim}$5--1$0{\times}$ more likely to be hosted in spirals ($\sim$40\%) or mergers ($\sim$20\%) than inactive (non-AGN) galaxies of the same stellar masses. 
\citet{ellison} did show however that {\em some} of the nuclear activity is triggered by interactions, showing that the AGN fraction increases by up to ${\sim}2.5{\times}$ for galaxies in close pairs with projected separations ${<}10$\,kpc, and that this enhancement in nuclear activity is greatest for equal-mass galaxy pairings. 
The morphological composition of our cluster X-ray AGN are broadly consistent with these previous studies, with 24/48 (50\%) being early-types (E/S0s), 8 undisturbed spirals (17\%), 9 disturbed spirals/mergers (19\%) and 7 unresolved objects (14\%), which are either low-luminosity galaxies or QSOs. We find a marginally higher fraction hosted by early-types, which could reflect the general increased prevalence of early-types among cluster galaxies, but could also be an overestimate due to our inability to robustly distinguish E/S0s and early-type spirals from the Subaru imaging. 

The frequent hosting of X-ray AGN (at least for $z{\la}1$) in otherwise undisturbed massive spirals appears inconsistent with the classical association between black hole accretion and bulge growth, and indeed the merger paradigm. As such we may have expected X-ray AGN to be preferentially hosted by early-type galaxies which have the most massive black holes, and hence must have had the most nuclear activity in the past. 
\citet{hop+hern} have developed a scenario for the fuelling of AGN in non-interacting spirals/S0s in which the accretion of cold gas onto supermassive black holes occurs via stochastic collisions with molecular clouds (or the inflow of these clouds via disk/bar instabilities). In these systems, the only requirement is the availability of cold gas within the disk, producing intermittent bursts of accretion at high Eddington ratios (${\sim}$1--10\%) and duty cycles of ${\sim}1$\%. The resultant feedback from the AGN is expected to have negligible impact on the host galaxy's disk and interstellar medium. 
The hosting of low-redshift X-ray AGN within massive spirals can thus be simply understood as the requirement of an available supply of cold gas. While elliptical galaxies host the most massive supermassive black holes, their frequent lack of cold gas means that there is no fuel for future gas accretion or black hole growth, and hence they are less likely to be X-ray luminous. \citet{kauffmann07} showed that early-type galaxies with strong nuclear activity (based on their O{\sc iii} emission) almost always have blue UV--optical colors and blue outer regions indicative of a reservoir of cold gas in the disk capable of feeding the AGN and star formation in the outer disk.

\section{Summary}
\label{sec:summary}

The key finding from our study of the distribution and host properties of 48 X-ray AGN identified from {\em Chandra} imaging of 26 massive clusters at $0.15{<}z{<}0.30$ is that they are clearly dynamically identified with an infalling population. This is manifest by their preferential location along the cluster caustics, complete avoidance of the caustic phase space with low relative velocities and cluster-centric radii, and their high velocity dispersion and non-Gaussian velocity distribution. The optically-selected Type I Seyferts/QSOs in our cluster sample show the same kinematical signatures. These provide the strongest observational constraints to date that the X-ray AGN and optically-selected Type I Seyferts/QSOs found in massive clusters are not a virialized population, and few if any can have resided within the dense ICM for a significant length of time. The cluster environment must very effectively suppress radiatively-efficient nuclear activity in its member galaxies. The dynamical properties of our X-ray AGN are very similar to those previously seen for late-type spiral galaxies in clusters, and a significant fraction ($35.4_{-7.5}^{+8.2}$\%) of the galaxies hosting the AGN are morphologically identified as spirals. This suggests that they are mostly triggered by secular processes such as bar/disk instabilities rather than mergers, although we do also find a number of X-ray AGN associated with ongoing mergers. 

These results, in conjunction with the previous clustering analyses of \citet{hickox} and \citet{starikova}, all indicate that two key requirements for radiatively-efficient nuclear activity to occur in galaxies is that they are the central galaxy within their dark matter halo, and have a ready supply of cold gas. This latter requirement is supported by the high fraction of X-ray AGN hosted by optically blue galaxies, both in clusters ($51{\pm}9$\%) and the general field at $z{\la}1$. 

\section*{Acknowledgments}

CPH and GPS acknowledge helpful discussions with Trevor Ponman and Kevin Pimbblet. CPH, GPS and AJRS acknowledge financial support from STFC.  GPS acknowledges support from the Royal Society. 
We acknowledge NASA funding for this project under the Spitzer program GO:40872. 
Observations reported here were obtained at the MMT Observatory, a joint facility of the University of Arizona and the Smithsonian Institution. The Millennium Simulation databases used in this paper and the web application providing online access to them were constructed as part of the activities of the German Astrophysical Virtual Observatory.

\clearpage
\begin{deluxetable}{lcclcrrrrll} 
\tablecolumns{11}
\tablewidth{0pt}
\tabletypesize{\scriptsize}
\tablecaption{\mbox{Catalog of spectroscopic X-ray AGN identified as cluster members.}}
\tablehead{ 
\vspace{0.15in}\\
\colhead{Cluster\hspace{-0.1in}} &  \colhead{RA} &  \colhead{Declination} &  \colhead{$z$} &  \colhead{$L_{X}$} &  \colhead{$v/\sigma_v$} & \colhead{$r/r_{500}$} &  \colhead{$M_{K}$} &  \colhead{SFR} &  \colhead{Morphology\hspace{-0.1in}} &  \colhead{Spectral} \\ 
\colhead{} & \colhead{(J2000)} & \colhead{(J2000)} & \colhead{} & \colhead{(erg/s)} & \colhead{} & \colhead{} & \colhead{(Vega)} & \colhead{\hspace{-0.1in}(M$_{\odot}$/yr)\hspace{-0.1in}} & \colhead{} & \colhead{Class}\\
}
\vspace{0.15in}\\
\startdata
A115  & 00:55:59.015 & +26:20:48.83 & 0.20293 &  $1.24{\times}10^{42}$ &  2.2355 &  0.0582 & -24.680 & 0.000 & E/S0 & AGN: H$\alpha$/NII  \\
A115  & 00:56:02.271 & +26:27:28.66 & 0.19036 &  $3.15{\times}10^{41}$ & -0.3121 &  0.5686 & -25.832 & 0.000 & E/S0 & AGN: H$\alpha$,NII \\ 
A115  & 00:56:08.656 & +26:25:10.87 & 0.19245 &  $3.28{\times}10^{41}$ &  0.1115 &  0.5996 & -25.317 & 0.986 & E/S0 & Passive \\ 
A115  & 00:56:11.427 & +26:27:37.24 & 0.19919 &  $1.77{\times}10^{42}$ &  1.4775 &  0.8145 & -23.504 & 5.157 & Unresolved & Broadline QSO \\
A115  & 00:56:19.702 & +26:21:53.24 & 0.19017 &  $9.12{\times}10^{41}$ & -0.3506 &  0.6700 & -25.895 & 1.022 & E/S0 & Passive\vspace{0.10in}\\ 
A209  & 01:31:33.789 & -13:32:23.14 & 0.21846 &  $1.58{\times}10^{42}$ &  1.7056 &  1.0846 & -24.549 & 5.442 & Sp/Int? & Star forming\\ 
A209  & 01:31:50.568 & -13:30:00.38 & 0.20255 &  $6.82{\times}10^{41}$ & -1.2249 &  1.1388 & -22.555 & 4.119 & Unresolved & Broad-line QSO \\
A209  & 01:32:16.328 & -13:35:38.15 & 0.20324 &  $1.17{\times}10^{42}$ & -1.0978 &  0.9475 & -23.104 & 4.407 & Merger & Noisy AGN: H$\alpha$,OII\vspace{0.10in}\\
A267  & 01:52:59.134 & +00:57:04.80 & 0.22773 &  $5.63{\times}10^{41}$ &  0.0542 &  1.2243 & -23.448 & 1.702 & Merger & Noisy AGN: H$\alpha$/NII\vspace{0.10in}\\ 
A383  & 02:48:24.679 & -03:31:45.56 & 0.18551 &  $1.05{\times}10^{42}$ & -1.0438 &  0.9561 & -24.747 & 1.117 & E/S0 & Passive\\ 
A383  & 02:48:21.892 & -03:34:25.55 & 0.186\tablenotemark{a} &  $3.16{\times}10^{42}$ & -0.8835 &  0.9608 & -25.188 & 1.125 & Merger & ------\vspace{0.10in} \\
A586  & 07:32:17.043 & +31:36:50.66 & 0.17378 &  $8.75{\times}10^{41}$ &  0.8855 &  0.1945 & -24.495 & 6.798 & Spiral* & Broadline QSO \\ 
A586  & 07:32:20.312 & +31:41:20.85 & 0.16583 &  $6.88{\times}10^{42}$ & -1.4001 &  0.5186 & -25.705 & 1.233 & E/S0 &  Passive\vspace{0.10in}\\ 
A611  & 08:00:56.472 & +36:05:25.02 & 0.27960 &  $9.67{\times}10^{41}$ & -1.6040 &  0.3828 & -24.929 & 4.406 & Sp/Int? & AGN \\
A611  & 08:00:52.901 & +36:06:26.11 & 0.27735 &  $6.88{\times}10^{41}$ & -2.1347 &  0.5939 & -25.781 & 0.000 & E/S0 & Passive\\ 
A611  & 08:01:02.229 & +36:02:45.73 & 0.27613 &  $5.27{\times}10^{42}$ & -2.4225 &  0.2384 & -23.336 & 4.743 & S0* & QSO \\
A611  & 08:01:06.125 & +36:03:16.83 & 0.27790 &  $6.67{\times}10^{41}$ & -2.0050 &  0.3560 & -24.978 & 18.295& Spiral* & AGN, OII,OIII\\ 
A611  & 08:01:04.616 & +36:02:21.21 & 0.27832 &  $1.14{\times}10^{42}$ & -1.9059 &  0.3564 & -23.486 & 0.000 & S0* & AGN? OII, H$\alpha$\vspace{0.10in} \\ 
A665  & 08:29:33.864 & +65:50:59.12 & 0.18795 &  $5.78{\times}10^{41}$ &  1.1041 &  1.1408 & -24.614 & 0.547 & Spiral & Just H$\alpha$\\
A665  & 08:30:41.665 & +65:58:10.68 & 0.18144 &  $7.99{\times}10^{41}$ & -0.2650 &  0.9509 & -25.987 & 0.000 & E/S0 & Broad H$\alpha$\\ 
A665  & 08:30:55.923 & +65:58:18.68 & 0.18477 &  $4.90{\times}10^{41}$ &  0.4353 &  0.9445 & -25.547 & 3.944 & Merger & Just H$\alpha$/NII\vspace{0.10in}\\ 
A963  & 10:17:00.712 & +39:04:32.76 & 0.21039 &  $7.62{\times}10^{43}$ &  1.3744 &  0.2863 & -25.077 & 46.893& E/S0 & QSO\vspace{0.05in} \\
A1689 & 13:11:16.981 & -01:16:56.63 & 0.1898\tablenotemark{b} &  $3.98{\times}10^{41}$ &  0.7731 &  0.5845 & -22.457 & 0.000 & E/S0 & ------\\ 
A1689 & 13:11:35.612 & -01:20:12.38 & 0.19955 &  $1.57{\times}10^{42}$ &  2.3768 &  0.1940 & -23.707 & 2.727 & Spiral* & Broadline QSO\\
A1689 & 13:11:22.161 & -01:23:45.68 & 0.1921\tablenotemark{b} &  $1.85{\times}10^{42}$ &  1.1514 &  0.4668 & -22.230 & 0.000 & S0* & ------\\ 
A1689 & 13:11:45.447 & -01:23:36.07 & 0.18685 &  $1.23{\times}10^{42}$ &  0.2878 &  0.6303 & -24.845 & 0.000 & E/S0 & Passive\vspace{0.10in}\\
A1758 & 13:32:44.206 & +50:31:07.82 & 0.28733 &  $1.32{\times}10^{42}$ &  1.3714 &  1.3013 & -22.824 & 4.088 & Unresolved*\hspace{-0.1in} & StarBurst\vspace{0.10in} \\
A1763 & 13:35:42.044 & +41:02:21.79 & 0.21816 &  $5.07{\times}10^{42}$ & -2.5409 &  0.9325 & -25.177 & 8.860 & Spiral & Broadline QSO\\
A1763 & 13:35:18.359 & +40:59:50.10 & 0.23972 &  $2.16{\times}10^{42}$ &  1.3333 &  0.0235 & -25.524 & 7.728 & dusty S0* & AGN: OIII,NII\vspace{0.10in}\\
A1835 & 14:00:51.309 & +02:59:05.52 & 0.25614 &  $1.99{\times}10^{42}$ &  0.6792 &  1.0289 & -24.176 & 4.308 & Unresolved & QSO \\
A1835 & 14:01:27.699 & +02:56:06.15 & 0.26472 &  $1.29{\times}10^{44}$ &  2.0868 &  1.0764 & -24.191 & 40.856& Unresolved & QSO\\
A1835 & 14:00:48.288 & +02:47:02.52 & 0.25640 &  $5.32{\times}10^{41}$ &  0.7218 &  0.9855 & -25.236 & 0.000 & E/S0 & Starburst\\ 
A1835 & 14:00:54.139 & +02:52:23.38 & 0.2617\tablenotemark{b} &  $3.26{\times}10^{42}$ &  1.5913 & 0.2980 & -22.810 & 0.000 & Unresolved & ------\\ 
A1835 & 14:01:08.836 & +02:44:57.67 & 0.2544\tablenotemark{b} &  $5.17{\times}10^{41}$ &  0.3937 &  1.1788 & -21.318 & 0.000 & Unresolved & ------\\ 
A1835 & 14:01:16.154 & +02:45:13.22 & 0.24468 &  $1.61{\times}10^{42}$ & -1.2009 &  1.2298 & -25.501 & 0.000 & E/S0 & Passive\vspace{0.10in}\\
A1914 & 14:25:46.628 & +37:49:24.40 & 0.17626 &  $8.85{\times}10^{41}$ &  2.1850 &  0.2861 & -22.887 & 0.000 & E/S0 & AGN: H${\alpha}$/NII\vspace{0.10in}\\ 
A2218 & 16:35:47.348 & +66:14:44.60 & 0.16855 &  $4.22{\times}10^{41}$ & -0.9628 &  0.2836 & -25.285 & 0.000 & E* & Passive\\
A2218 & 16:35:56.090 & +66:16:15.13 & 0.17932 &  $1.94{\times}10^{41}$ &  1.2202 &  0.5043 & -22.359 & 12.967& S0* & Noisy AGN\\ 
A2218 & 16:36:30.913 & +66:15:05.24 & 0.18172 &  $3.75{\times}10^{42}$ &  1.7067 &  0.6773 & -23.671 & 1.036 & Spiral*& Broadline AGN\vspace{0.10in}\\
A2219 & 16:40:05.839 & +46:43:41.43 & 0.22891 &  $3.91{\times}10^{42}$ &  0.5688 &  0.4599 & -23.024 & 2.015 & E/S0 & AGN\\ 
A2219 & 16:40:09.337 & +46:44:19.55 & 0.22993 &  $1.76{\times}10^{42}$ &  0.7495 &  0.4367 & -24.572 & 48.478& Sp/Int? & StarBurst\vspace{0.10in}\\
A2390 & 21:53:31.422 & +17:41:33.68 & 0.24587 &  $1.03{\times}10^{42}$ &  2.9608 &  0.1902 & -24.577 & 5.658 & Spiral* & QSO \\ 
A2390 & 21:53:25.330 & +17:43:21.75 & 0.2242\tablenotemark{c} &  $1.92{\times}10^{42}$ & -0.8424 &  0.4667 & -23.727 & 0.000 & E/S0 & ------\\
A2390 & 21:53:37.972 & +17:43:47.46 & 0.24071 &  $6.44{\times}10^{41}$ &  2.0552 &  0.3044 & -23.289 & 14.819& S0* & QSO \\ 
A2390 & 21:53:45.559 & +17:41:47.88 & 0.21437 &  $1.87{\times}10^{42}$ & -2.5677 &  0.3042 & -25.026 & 52.559& Sp/Int? & QSO \\ 
A2390 & 21:53:45.738 & +17:41:08.46 & 0.23483 &  $6.22{\times}10^{41}$ &  1.0232 &  0.3219 & -23.755 & 0.000 & E/S0 & Passive\vspace{0.10in} \\ 
A2485 & 22:48:30.816 & -16:10:30.94 & 0.24139 &  $1.84{\times}10^{43}$ & -1.4940 &  1.1466 & -25.419 & 39.317& Sp/Int? & Broadline AGN\vspace{0.10in}\\
Z1883 & 08:42:51.963 & +29:28:25.02 & 0.18442 &  $1.22{\times}10^{42}$ & -2.7000 &  0.2315 & -23.848 & 99.964& S0* & QSO\\ 
\vspace{0.1in}
\enddata
\tablecomments{Col. 1: Cluster Name. Cols. 2 \& 3: The right ascension and declination of the X-ray AGN (J2000). Col. 4: Redshift. Col. 5: Broadband (0.3--7\,keV) X-ray luminosity of X-ray AGN assuming $\Gamma{=}1.7$ power law spectrum. Col. 6: Velocity offset relative to the cluster redshift in units of $\sigma_v$, the velocity dispersion of the cluster. Col. 7: Projected cluster-centric radius in units of $r_{500}$. Col. 8: $K$-band absolute magnitude of host galaxy (Vega). Col. 9: SFR of host galaxy in units of M$_{\odot}$yr$^{-1}$. The SFR is derived from the 24$\mu$m luminosity, and is set to zero for {\em Spitzer} non-detections. Col. 10: Morphological class of host galaxy. E/S0: Early-type galaxy. Sp/Int?: Spirals with signs of possible interations. Starred morphological classifications indicate those host galaxies for which HST imaging was available. Col. 11: Spectroscopic class derived from our MMT/Hectospec data where available.}
\tablenotetext{a}{Redshift from \citet{rizza}.}
\tablenotetext{b}{Redshifts from \citet{czoske}.}
\tablenotetext{c}{Redshift from \citet{crawford}.}
\label{catalog}
\end{deluxetable}

\label{lastpage}
\end{document}